\documentclass[twocol]{ametsocV6.1}
\usepackage{physics}
\usepackage{graphicx,nicefrac}
\usepackage{adjustbox}
\usepackage{float}

\newcommand*{\vect}[1]{\mathbf{#1}}

\title{PQG-DL-Ekman: a triple-deck boundary layer theory for large-scale atmospheric flow with moist process closures}
\authors{Daniel B\"{a}umer,\aff{a}\correspondingauthor{Daniel B\"{a}umer, daniel.baeumer@univie.ac.at} 
Rupert Klein\aff{b} 
}

\affiliation{\aff{a}{Research platform MMM, c/o Fakult\"{a}t f\"{u}r Mathematik, Universit\"{a}t Wien, Vienna, Austria} \\
\aff{b}{FB Mathematik und Informatik, Freie Universit\"{a}t Berlin, Berlin,
Germany}
}

\abstract{Reduced mathematical models for atmospheric dynamics at various scales have a long and rich history. However, versions of such models that explicitly incorporate moisture and phase changes have been developed only fairly recently. This work merges one of said modeling innovations, namely Smith and Stechmann's \emph{precipitating quasigeostrophic} (PQG) model family, with a triple-deck boundary layer theory due to Klein et al.~that extends the classical QG-Ekman theory by an intermediate \emph{diabatic layer} (DL). A detailed asymptotic analysis of the Clausius-Clapeyron relation and Kessler-type bulk microphysics closures is included in the systematic derivation of the resulting PQG-DL-Ekman theory. Furthermore, to illustrate some of the model's properties, explicit axisymmetric solutions of the precipitating diabatic layer equations are derived and combined with numerical sample solutions for the bulk flow.}

\begin{document}

\maketitle

\section{Introduction}
Moist processes, from nucleation on the smallest cloud condensation kernels to large-scale precipitation systems, play a pivotal role in the dynamics of our planet's atmosphere. In the theoretical treatment of said dynamics, scale analysis, boundary layer theory and, more recently, the method of multiple scales have played a central role for many decades, leading to reduced models such as the ubiquitous quasi-geostrophic (QG) theory for the midlatitudes, the Matsuno-Gill models for the tropics and many more. The explicit incorporation of the most important moist processes, such as cloud formation and rain production, as well as the associated cloud microphysics, in these models thus promises to further our understanding of the highly complex interactions between the bulk flow and its (moist) thermodynamics at various scales - yet, first steps in this direction have been taken only relatively recently: the first multiscale model that systematically investigated transport equations for moisture species \emph{and} the corresponding bulk microphysics closures in the context of asymptotic analysis was proposed by \citet{klein2006}. Numerous studies on moist convection by Majda and coauthors followed, such as \citet{majda2007}, which pursued systematic strategies for superparameterization by multiple scales asymptotics, or \citep{hernandez2013}, where a cloud-resolving model was proposed that later served as the starting point for the derivation of the original precipitating QG equations - one of the building blocks for the present work. The dynamics of convective cloud towers in the tropics was further studied by \citet{hittmeir2018}.

Regarding synoptic-scale flow in the midlatitudes, the first systematic extension of classical QG theory by transport equations for water vapor and precipitation is due to \citet{smith2017}, who derived the family of \emph{precipitating QG} (PQG) models in a manner that was consistent with the textbook treatment of dry QG as in, e.g., \citep{pedlosky1987}. This work provided the foundation for many follow-up studies on topics such as PQG's nonlinear potential vorticity inversion and the study of balanced and unbalanced components of moist atmospheric flow \citep{wetzel2019,wetzel2020}. 

A different ansatz aiming at improvements over the classical QG-Ekman theory \citep{pedlosky1987,vallis2017} was pursued by \citet{klein2022}: in their work, the observation that diabatic processes tend to be systematically stronger at lower tropospheric levels, more frequently allowing for neutral or even unstable stratification, led to the extension of the QG-Ekman model to a \emph{triple-deck} boundary layer theory. This theory augments QG-Ekman with the novel \emph{diabatic layer} (DL) as an intermediary between the free troposphere, which is governed by QG, and the (frictional) planetary boundary layer, where Ekman applies. While also applicable to a dry atmosphere, this modeling innovation was motivated in large part by empirical evidence that the intensity of processes such as thermally driven convection or cooling by evaporation of rain at lower tropospheric levels often significantly exceed the strength of heat sources that the scaling assumptions of QG theory permit. By design, such a triple-deck theory is expected to explain interesting atmospheric phenomena with strong lower atmospheric diabatic effects. Indeed, \citet{klein2022} demonstrate that a QG-DL-Ekman solution representing an axisymmetric convection-driven vortex reproduces important features of ``heat lows'' as discussed, e.g., by \citet{RaczSmith1999}. Similar effects, but now based on evaporative cooling, may be expected when rain in midlatitude weather precipitates into under-saturated air below cloud bottom as discussed below.

Therefore, it should be a worthwhile endeavor to combine the modeling efforts of \citet{smith2017} and \citet{klein2022} and to derive a refined \emph{PQG-DL-Ekman} model that takes into account the varying strengths of both spatiotemporal moisture perturbations and their associated phase changes across the various tropospheric layers. The foundation for this undertaking was laid by \citet{baeumer2023}, where the authors first showed how Smith and Stechmann's PQG equations can be embedded into the framework of \citet{klein2010} for model hierarchies in atmospheric flows, and subsequently derived a new variant of PQG that was designed specifically to connect to a diabatic layer with matching moist process closures.

In the present work, we build on said foundation and derive the previously announced PQG-DL-Ekman model, strictly in keeping with the tenets of singular perturbation theory \citep{kevorkian1996}. By integrating bulk microphysics closures systematically into such a scale-specific model we hope to advance our understanding of the respective strengths and influences of moist processes at the scales under consideration. More specifically, just like classical QG(-Ekman) theory captures the essential dynamical features of extratropical cyclones, this extended theory should paint a clearer picture of cloud formation and precipitation in these weather phenomena, as well as the interactions of these processes with the large-scale flow. In the final section, some of the model's properties are illustrated by numerical simulations of cloud formation triggered by an idealized (axisymmetric) low-level vortex.
\section{The governing equations}
Building upon \citep{hittmeir2018,baeumer2023}, we adopt their formulation of the full compressible flow equations including established bulk microphysics closures \citep{kessler1995,grabowski1996,klein2006} and supplemented by \emph{a priori} unspecified closures for turbulent friction and diffusion. For its fully viscous--diffusive version, local wellposedness has  been established by \citet{doppler2024}:
\begin{subequations}\label{eq:GoverningEquations}
\begin{align}
\label{eq:GoverningEquations-a}
D_t\vect{u}+2(\vect{\Omega}\times\vect{v})_\parallel+\frac{1}{\rho}\grad_\parallel{p}
  = & \frac{\rho_d}{\rho}q_r V_r\partial_z\vect{u}+\mathcal{D}_{\vect{u}}, 
    \\
\label{eq:GoverningEquations-b}
D_tw+2(\vect{\Omega}\times\vect{v})_\perp+\frac{1}{\rho}\partial_zp
  = & -g+\frac{\rho_d}{\rho}q_r V_r\partial_z w+\mathcal{D}_w, 
    \\
\label{eq:GoverningEquations-c}
D_t\rho_d+\rho_d(\div{\vect{v}})
  = & 0, 
    \\
\label{eq:GoverningEquations-d}
\hspace{-10pt}
C D_t\ln\left(\frac{\theta}{\theta_{\text{ref}}}\right)+\Sigma D_t\ln\left(\frac{p}{p_{\text{ref}}}\right) & \nonumber \\
+\frac{L(T)}{T}D_tq_v = & c_lV_rq_r\left(\partial_z\ln\left(\frac{\theta}{\theta_{\text{ref}}}\right)\right. \nonumber \\
&\left.+\frac{R_d}{c_{\text{pd}}}\partial_z\ln\left(\frac{p}{p_{\text{ref}}}\right)\right) + Q + \mathcal{D}_\theta, 
    \\
\label{eq:GoverningEquations-e}
D_tq_v
  = & S_{\text{ev}}-S_{\text{cd}}+\mathcal{D}_v, 
    \\
\label{eq:GoverningEquations-f}
D_tq_c
  = & S_{\text{cd}}-S_{\text{cr}}-S_{\text{ac}}+\mathcal{D}_c, 
    \\
\label{eq:GoverningEquations-g}
D_tq_r-\frac{1}{\rho_d}\partial_z(\rho_dV_rq_r)
  = & S_{\text{cr}}+S_{\text{ac}}-S_{\text{ev}}+\mathcal{D}_r,
\end{align}
\end{subequations}
with the additional relations
\begin{subequations}\label{eq:GoverningEquationsSupplement}
\begin{align}
\label{eq:GoverningEquationsSupplement-a}
\rho
  & =\rho_d(1+q_v+q_c+q_r), 
    \\
\label{eq:GoverningEquationsSupplement-b}
p
  & = R_d\rho_dT(1+\frac{q_v}{R_d/R_v}), 
    \\
\label{eq:GoverningEquationsSupplement-c}
T
  & = \theta \left(\frac{p}{p_{\text{ref}}}\right)^{\frac{\gamma-1}{\gamma}} \equiv \theta \pi, 
    \\
\label{eq:GoverningEquationsSupplement-d}
\vect{v}
  & = \vect{u}+w\vect{k}, 
    \\
\label{eq:GoverningEquationsSupplement-e}
S_{\text{ev}}
  & = C_{\text{ev}}\frac{p}{\rho}(q_{\text{vs}}-q_v)^+q_r, 
    \\
\label{eq:GoverningEquationsSupplement-f}
S_{\text{cd}}
  & = C_{\text{cn}}(q_v-q_{\text{vs}})^+q_{\text{cn}}+C_{\text{cd}}(q_v-q_{\text{vs}})q_c, 
    \\
\label{eq:GoverningEquationsSupplement-g}
S_{\text{ac}}
  & = C_{\text{ac}}(q_c-q_{\text{ac}})^+, 
    \\
\label{eq:GoverningEquationsSupplement-h}
S_{\text{cr}}
  & = C_{\text{cr}}q_cq_r, 
    \\
\label{eq:GoverningEquationsSupplement-i}
C 
  & = c_{\text{pd}}+c_{\text{pv}}q_v+c_l(q_c+q_r), 
    \\
\label{eq:GoverningEquationsSupplement-j}
\Sigma
  & = (\frac{c_{\text{pv}}}{c_{\text{pd}}}R_d-R_v)q_v+\frac{c_l}{c_{\text{pd}}}R_d(q_c+q_r),
    \\
\label{eq:GoverningEquationsSupplement-k}
L(T) 
  & = L_{\text{ref}}-(c_l-c_{\text{pv}})(T-T_{\text{ref}})\equiv L_{\text{ref}}\phi(T).
\end{align}
\end{subequations}
The saturation vapor content, $q_{\text{vs}}$, is defined via the ideal gas Clausius-Clapeyron (CC) relation
\begin{equation}\label{CC_diff_qvs_formula}
    \frac{d \ln e_s}{dT}=\frac{L(T)}{R_vT^2}
    \qquad\text{and}\qquad
    q_{\text{vs}}=\frac{R_d}{R_v}\frac{e_s}{p-e_s}.
\end{equation}
In summary, \\ $(\vect{u}=(u,v,0),w,\rho,\rho_d,T,\theta,p,\pi,e_s,q_v,q_c,q_r,q_{\text{vs}})$ denote horizontal and vertical velocity, density, dry air density, temperature, potential temperature, pressure, the dimensionless Exner pressure, the saturation vapor pressure, and the mixing ratios (i.e.~ratios of the respective densities to the \emph{dry} air density) of water vapor, cloud water and rain, and the saturation mixing ratio, respectively; $g \approx 9.8\;\text{m s}^{-2}$ is the gravitational acceleration, $\vect{\Omega}$ the earth rotation vector, and the subscripts $\parallel$ and $\perp$ indicate horizontal and vertical components of a vector, respectively. The positive part of a function $f$ is $f^+$; $c_{\text{pd}}$ and $c_{\text{pv}}$ are the specific heat capacities at constant pressure of dry air and water vapor, $c_l$ is the heat capacity of liquid water, assumed constant for simplicity, and $V_r$ is the terminal rainfall velocity; $R_d$ and $R_v$ are the gas constants for dry air and water vapor, $\gamma=c_{\text{pd}}\slash(c_{\text{pd}}-R_d)$ is the isentropic exponent of dry air, $\vect{k}$ the vertical unit vector, $p_{\text{ref}} \approx 10^5\text{ Pa}$ the reference pressure and the material derivative $D_t$ is given by
\begin{equation}
D_t=\partial_t+\vect{v}\cdot\grad=\partial_t+\vect{u}\cdot\grad_\parallel+w\partial_z.
\end{equation}
In line with \citep{cotton2011}, the dry air density, $\rho_d$, obeys the continuity equation~\eqref{eq:GoverningEquations-c}. Notice that $\rho$, in \eqref{eq:GoverningEquations-a}, \eqref{eq:GoverningEquations-b}, \eqref{eq:GoverningEquationsSupplement-e}, is the full density from \eqref{eq:GoverningEquationsSupplement-a} and not the dry air density. Therefore, the effect of moisture on (total) density is properly accounted for. Individual contributions from the moist constituents are the following: in the momentum equations \eqref{eq:GoverningEquations-a}-\eqref{eq:GoverningEquations-b}, momentum changes due to falling rain appear on the right-hand side; in the thermodynamic equation \eqref{eq:GoverningEquations-d}, $C$ denotes the ``total moist heat capacity'', specified in \eqref{eq:GoverningEquationsSupplement-i}; $\Sigma$, defined in (2j), collects moist contributions related to the work done by pressure forces and $L(T)$ is the latent heat (enthalpy) of vaporization, which can be written as a linear function of temperature under the assumption of constant $c_l$.  The reference values $L_{\text{ref}}$ for latent heat and $T_{\text{ref}}$ for temperature are listed in Table \ref{tab:moist_thermodynamics}, while $Q$ represents all diabatic heat sources other than latent heating. Finally, the right-hand side represents temperature changes due to precipitation falling relative to its environment. In the transport equations for the respective mixing ratios \eqref{eq:GoverningEquations-e}-\eqref{eq:GoverningEquations-g}, terms of the form $S_\star$ denote Kessler-type closures for the microphysical processes of evaporation (ev), condensation (cd), autoconversion (ac) and collection (cr), respectively. Here, all phase transitions are functions of the respective mixing ratios only. For simplicity in the theoretical and numerical treatment, these closures have been implemented in (bi)linearized form, as in \citep{hittmeir2018}. In the corresponding Eqs.~\eqref{eq:GoverningEquationsSupplement-e}-\eqref{eq:GoverningEquationsSupplement-h},parameters of the form $C_\star$ denote rate constants of the respective processes, while $q_{\text{cn}}$ represents the local density of condensation kernels and $q_{\text{ac}}$ denotes an activation threshold for the autoconversion of cloud droplets into raindrops. We emphasize that, while the phase transitions themselves are modeled in fairly simple fashion, there is a nonlinear two-way coupling between the moist constituents and the (thermo)dynamics of the flow in this formulation, primarily via moist contributions to the thermodynamic equation \eqref{eq:GoverningEquations-d} on the one hand, and the transport operator in \eqref{eq:GoverningEquations-e}-\eqref{eq:GoverningEquations-g}, as well as the CC relation \eqref{CC_diff_qvs_formula} on the other hand.

Turbulent closures for the quantity $\star$ are indicated by $\mathcal{D}_\star$. We defer a discussion of their respective importance and explicit formulas to subsection~3.c.

We do not explicitly consider cold (cirrus) clouds that would necessitate parametrization of the ice phase. While the importance of ice formation and the associated phase changes in the formation of stratiform cloud decks is undisputed \citep{houze2014}, their systematic incorporation would give rise to a host of additional modeling difficulties. This endeavor is deferred to future work.
\begin{table}[H]
\centering
\begin{adjustbox}{width=0.48\textwidth}
\begin{tabular}{l l l l}
\hline
    $T_{\text{ref}} = \theta_{\text{ref}}$ & 288.15 & K & Typical midlatitude surface temperature \\
    $R_d$ & 287 & J/kg/K & Dry air gas constant \\
    $c_{\text{pd}}$ & 1005 & J/kg/K & Specific heat capacity of dry air at constant pressure \\
    $R_v$ & 462 & J/kg/K & Water vapor gas constant \\
    $c_{\text{pv}}$ & 1850 & J/kg/K & Specific heat capacity of water vapor at constant pressure \\
    $c_l$ & 4186 & J/kg/K & Specific heat capacity of liquid water \\
    $L_{\text{ref}}$ & $2.466 \times 10^6$ & J/kg & Latent heat (enthalpy) of vaporization at $T=T_{\text{ref}}$ \\
    \hline
\end{tabular}
\end{adjustbox}
\caption{Thermodynamic state parameters for a mixture of dry air, water vapor and liquid water.}
\label{tab:moist_thermodynamics}
\end{table}
\section{Scaling: non-dimensionalization and choice of a distinguished limit}
\label{sec:DistinguishedLimit}
As in \citep{baeumer2023}, we base our derivation on the distinguished limit for dry airflows of \citet{klein2010}: with $\epsilon$ representing a generic small dimensionless parameter, with a value of $\sim0.1$ for Earth's atmosphere, we scale the Mach, external Froude and Rossby numbers as
\begin{align}
&\text{M} 
= \frac{u_{\text{ref}}}{\sqrt{\nicefrac{p_{\text{ref}}}{\rho_{\text{ref}}}}}=\epsilon^{\nicefrac{3}{2}}
= \frac{u_{\text{ref}}}{\sqrt{gh_{\text{sc}}}}=\text{Fr}_\text{ext}, \nonumber \\
&\text{Ro} = \frac{u_{\text{ref}}}{2\Omega l_{\text{ref}}}=O(\epsilon),
\end{align}
where the relevant parameters are summarized in Table \ref{tab:synoptic_scale}. With $l_{\text{ref}}$ the synoptic scale, we have
\begin{equation}
    \frac{h_{\text{sc}}}{l_{\text{ref}}}=\epsilon^2
\end{equation}
as our scaling for the aspect ratio of the flow, which suggests a reference vertical velocity of
\begin{equation}
    w_{\text{ref}} = \epsilon^2 u_{\text{ref}}.
\end{equation}
\begin{table}[H]
    \centering
    \begin{adjustbox}{width=0.48\textwidth}
    \begin{tabular}{l l l l}
    \hline
       $u_{\text{ref}}$  & $\sim 10$ & m $\text{s}^{-1}$ & Typical horizontal wind speed \\
       $l_{\text{ref}}$  & $\sim 10^6$ & m & Synoptic length scale \\
       $h_{\text{sc}}=\frac{R_d T_{\text{ref}}}{g}$ & $\sim 8.5 \times 10^3$ & m & Pressure scale height \\
       $w_{\text{ref}}=u_{\text{ref}} \frac{h_{\text{sc}}}{l_{\text{ref}}}$ & $\sim 0.1$ & m $\text{s}^{-1}$ & Reference vertical velocity \\
       \hline
    \end{tabular}
    \end{adjustbox}
    \caption{Synoptic reference lengths and velocities.}
    \label{tab:synoptic_scale}
\end{table}
%
%In our treatment of the Coriolis force, we employ 
Adopting the $\beta$-plane approximation \citep{pedlosky1987} with reference latitude $\sim45^{\circ}$, we have
\begin{equation}
    \frac{1}{\Omega}\vect{\Omega}\sim(f_0+\epsilon\beta\frac{y}{l_{\text{ref}}})\vect{k}+o(\epsilon).
\end{equation}
Next, we supplement our distinguished limit with scalings for the thermodynamic parameters of moist, cloudy air. In doing so, we consider the two options first identified by \citet{hittmeir2018} as viable choices, shown in Table \ref{tab:regimes_moist} below:
\begin{table}[H]
    \centering
    \begin{adjustbox}{width=0.48\textwidth}
        \begin{tabular}{l l l l}
    \hline
         & Value & Regime $\alpha = 1$ & Regime $\alpha = 0$ \\
         \hline
        \emph{Nondimensional parameters} & & & \\
        $\frac{R_d}{c_{\text{pd}}}$ & $0.29$ & $\epsilon \Gamma$ & $\epsilon \Gamma$ \\
        $\frac{c_{\text{pv}}}{c_{\text{pd}}}$ & $1.8$ & $k_v$ & $\epsilon^{-1}k_v$ \\
        $\frac{R_v}{c_{\text{pd}}}$ & $0.46$ & $1/A$ & $1/A$ \\
        $\frac{c_l}{c_{\text{pd}}}$ & $4.2$ & $\epsilon^{-1}k_l$ & $\epsilon^{-1}k_l$ \\
        $\frac{L_{\text{ref}}}{c_{\text{pd}}T_{\text{ref}}}$ & $9.1$ & $\epsilon^{-1}L$ & $\epsilon^{-1}L$ \\
        \emph{Derived nondimensional parameters} & & & \\
        $\frac{R_d}{R_v}$ & $0.62$ & $E$ & $\epsilon E$ \\
        $\frac{c_{\text{pv}}}{c_{\text{pd}}}\frac{R_d}{c_{\text{pd}}}-\frac{R_v}{c_{\text{pd}}}$ & $0.067$ & $\epsilon \kappa_v$ & $\kappa_v$ \\
        \hline
    \end{tabular}
    \end{adjustbox}
    \caption{Two viable moist extensions of the distinguished limit for dry air}
    \label{tab:regimes_moist}
\end{table}
In the two right-most columns, all parameters are bounded \emph{independent} of $\epsilon$ - in other words, they are $O(1)$ in the limit $\epsilon\rightarrow0$. Let us now focus on the regime labeled $\alpha = 1$, which assigns asymptotic rescalings based on the numerical magnitudes of the respective parameters. Here, we are faced with the following dilemma: if we were to adhere to the tenets of similarity theory, the distinguished limit defined by the top five rows would imply that the derived quantity
\begin{equation} \label{scaling_moist_par_1}
    \frac{R_d}{R_v}=\frac{R_d}{c_{\text{pd}}}/\frac{R_v}{c_{\text{pd}}}=\epsilon\Gamma A
\end{equation}
is asymptotically small, while
\begin{equation}
    \frac{c_{\text{pv}}}{c_{\text{pd}}}\frac{R_d}{c_{\text{pd}}}-\frac{R_v}{c_{\text{pd}}}=k_v\epsilon\Gamma-\frac{1}{A}
\end{equation}
attains a negative value in the limit $\epsilon \rightarrow 0$. However, $R_d / R_v$ is actually \emph{larger} than $R_v / c_{\text{pd}}$, which we assigned a bounded value, while the derived expression at the bottom of Table \ref{tab:regimes_moist} is a small \emph{positive} quantity. In proposing the value-based regime $\alpha = 1$, we therefore take the following position: \emph{accurate representation of the numerical values takes precedence over strict formal consistency,} and the ad hoc scalings for the two derived quantities are thus viewed as justified. From the perspective of dimensional analysis, the resulting formal inconsistency is of course worrisome, but the dynamics obtained in this fashion extend the classical QG theory in a very natural way and are compatible with the PQG model family. We therefore adopt said regime {$\alpha = 1$, also considered by \citet{hittmeir2018},} as the basis for the present investigations. Consequences of the formally consistent regime, $\alpha = 0$, are briefly addressed in section~7, while the review article by \citet{baeumer2025} provides a broader discussion of viable distinguished limits for moist air.
\subsection{The Clausius-Clapeyron relation}
\label{ssec:ClausiusClapeyron}
We recall from \eqref{CC_diff_qvs_formula} the Clausius-Clapeyron relation for an ideal gas,
\begin{equation}\label{CC_diff}
    \frac{d \ln e_s}{dT}=\frac{L(T)}{R_vT^2},
\end{equation}
relating the saturation vapor pressure, $e_s$, to temperature, $T$. With the linearization \eqref{eq:GoverningEquationsSupplement-k} of latent heat, i.e, $L(T)=L_{\text{ref}}-(c_l-c_{\text{pv}})(T-T_{\text{ref}})$ with $c_l, c_{pv} = \text{const.}$, \eqref{CC_diff} can be integrated exactly,
\begin{align}\label{es_exact}
    e_s(T)=&{e_s}_{\text{ref}}\left(\frac{T_{\text{ref}}}{T}\right)^{\frac{c_l-c_{\text{pv}}}{R_v}}\nonumber \\ &\times\exp{\left(\frac{L_{\text{ref}}}{R_vT_{\text{ref}}}+\frac{c_l-c_{\text{pv}}}{R_v}\right)\left(1-\frac{T_{\text{ref}}}{T}\right)}.
\end{align}
As shown in section 4, the chosen scalings for the various thermodynamic parameters, notably the so-called Newtonian limit $R_d/c_{pd} = \epsilon\Gamma$ \citep{ParkinsEtAl2000,klein2006,baeumer2025}, lead to an exponent that is \emph{bounded} across the whole troposphere as $\epsilon\to0$. Finally, the typical midlatitude value of ${e_s}_{\text{ref}}=e_s(T_{\text{ref}})\approx 1704 \text{ Pa}$ suggests the scaling
\begin{equation}
    \frac{{e_s}_{\text{ref}}}{p_{\text{ref}}}=O(\epsilon^2)\,.
\end{equation}
\subsection{Moist constituents and phase changes}
Here we discuss the asymptotics scalings of the terminal rain fall velocity, the water vapor, cloud water, and rain water content, and of the conversion terms in the respective balances~\eqref{eq:GoverningEquations-e}-\eqref{eq:GoverningEquations-g}.
\subsubsection{Terminal rainfall velocity}
The terminal velocity $V_r$ of raindrops depends on their size as  attained during their passage through the clouds. Models of large-scale dynamics, in which \emph{all} hydrometeors with significant fall speed are lumped into one variable (here: $q_r$), cannot account for this dependence. Thus, one assumes $V_r = \text{const.}$, to be understood as weighted average of terminal velocities over the size spectrum. \citet{baeumer2023}, citing \citep{houze2014}, argue that the stratiform precipitation predominantly found in midlatitude cyclones tends to be gentler and produces smaller raindrops than convective precipitation and conclude that in a QG-based model, $V_r$ should not exceed a few meters per second on average. This assumption is also supported by observational data, see, e.g., \citep{niu2010}. The appropriate scaling then reads
\begin{equation}
    \frac{V_r}{w_{\text{ref}}}=\epsilon^{-1}V_T
    \qquad\text{or}\qquad
    \frac{V_r}{u_{\text{ref}}}=\epsilon V_T
\end{equation}
with $V_T$ bounded independent of $\epsilon$. In contrast, \citet{hittmeir2018} adopt $V_r \sim u_{\text{ref}}$ for (tropical) convective clouds. In a future multiscale regime, one might therefore need to incorporate the dependence of the average fall speed on local atmospheric conditions.
\subsubsection{Water vapor}
Since the saturation mixing ratio, $q_{\text{vs}}$, and the saturation vapor pressure, $e_s$, are related by $q_{\text{vs}}=\frac{R_d}{R_v}\frac{e_s}{p-e_s}$ (see \eqref{CC_diff_qvs_formula}), the scalings in section~\ref{sec:DistinguishedLimit}.\ref{ssec:ClausiusClapeyron} imply
\begin{equation}\label{q_vs_scaled}
    q_{\text{vs}}=O(\epsilon^2).
\end{equation}
From the Clausius-Clapeyron relation, we obtained \eqref{es_exact} for $e_s(T)$. In the QG and DL regimes discussed here, pressure and temperature vary about fixed vertical background distributions with small amplitude. As a consequence, $q_{\text{vs}}$ varies only in the vertical at leading order as well. It thus remains to determine appropriate scalings for deviations of the actual water vapor content about this ``moist background''. Here, we should keep the following in mind:
\begin{enumerate}
    \item Relative supersaturation, $\frac{q_\text{v} - q_{\text{vs}}}{q_{\text{vs}}} > 0$, hardly exceeds $1\%$ \citep{houze2014}. This is reflected and guaranteed here by an according stiffness of the condensation term in \eqref{eq:GoverningEquationsSupplement-f}.
    \item \label{subsat_large} In contrast, subsaturation is often substantial with, e.g., surface relative humidities in Germany averaging to $78.6\%$ between 1971 and 2000 \citep{razafimaharo2020}. Also, relative humidity typically decreases further with height.
\end{enumerate}
Guided by item \ref{subsat_large}., \citet{baeumer2023} assumed variations of $q_v$ unrestricted by a moist background for their $\text{PQG}_{\text{DL}}$ model, permitting significant subsaturation. Due to the scaling assumptions of QG theory, however, this necessitated the scaling $q_v\sim q_{\text{vs}}\sim\epsilon^3$ for the bulk troposphere. Since this is not realistic close to the surface, it required an increase in asymptotic magnitude of $q_{\text{vs}}$ in the DL and Ekman layers. Closer inspection revealed that such a behavior of $q_v\sim q_{\text{vs}}$ would not be compatible with Clausius-Clapeyron unless the entire approach to the vertical layering of the atmosphere was changed altogether. This, while theoretically viable, would have forced a different distinguished limit rendering further comparisons with previously derived models difficult -- in short, it would have been rather impractical. Therefore, we decided to keep the scaling \eqref{q_vs_scaled} for the saturation mixing ratio here and to set
\begin{equation} \label{qv_background}
    q_v = q_{\text{vs}} + o(1)
\end{equation}
for the scaled variables. With this assumption, the model is restricted to atmospheric regions with consistently high humidity, similar to the original PQG model of \citet{smith2017}. For the perturbations
\begin{equation}
    \tilde{q}_v:=q_v-q_{\text{vs}},
\end{equation}
we based our scaling on the following observations:
\begin{enumerate}
     \item To retain a nontrivial influence of moist processes, the balance equations for water vapor and potential temperature in  \eqref{eq:GoverningEquations-d} need to be coupled at leading order.
   \item \citet{houze2014} identifies \emph{warm frontal lifting} as the main synoptic scale source of liquid water: Moist air moving upwards along a gentle slope produces clouds by continually lifting air parcels across their lifting condensation levels. Thus, the leading-order transport of water vapor should incorporate \emph{both} horizontal and vertical advection in the bulk troposphere.
\end{enumerate}
Recalling the respective strengths of potential temperature perturbations in QG and DL, we conclude that the only scaling in agreement with both these observations is
\begin{subequations}
    \begin{align}
        \tilde{q}_v^{\text{QG}}&=O(\epsilon^3)\quad \text{and} \label{qv_scaled_qg}\\
        \tilde{q}_v^{\text{DL}}&=O(\epsilon^{5/2}). \label{qv_scaled_dl}
    \end{align}
\end{subequations}
\subsubsection{Cloud liquid water}
In asymptotic models involving clouds and precipitation, the cloud liquid water content (CLWC) is commonly assumed small compared to the water vapor content in line with observations \citep{zhao2014, zhang2020}. Here, we assume accordingly that
\begin{equation}
    q_c^{\text{QG}}=O(\epsilon^3)\,.
\end{equation}
For the diabatic layer, the scaling in \eqref{qv_scaled_dl} seems to imply a progression toward significantly \emph{higher} CLWCs, i.e., $q_c\sim\epsilon^{5/2}$. This would correspond, however, to mixing ratios of several g/kg, which is not supported by observations. Hence, we keep the QG scaling and assume
\begin{equation}\label{qc_scaled_dl}
    q_c^{\text{DL}}=O(\epsilon^3).
\end{equation}
As shown in section~4, this scaling eliminates cloud water \emph{transport} from the leading-order dynamics in DL, relegating the \emph{initial formation} of clouds to the middle and upper troposphere.
\subsubsection{Rain}
Stratiform precipitation, the dominant form of precipitation in midlatitude cyclones, is markedly different from convective precipitation \citep{houze2014}: updrafts are weaker, raindrops remain smaller and thus exhibit slower average fall speeds as discussed above. Accordingly, the fraction of cloud droplets growing to precipitable size is smaller than it is in convective systems.

Following arguments of \citep{baeumer2023} we adopt their QG-scaling
\begin{equation}\label{qr_scaled_pqg}
    q_r^{\text{QG}}=O(\epsilon^4),
\end{equation}
of the rain amount. Yet, in contrast to their work, we maintain this magnitude for the DL, i.e.,
\begin{equation}\label{qr_scaled_pdl}
    q_r^{\text{DL}}=O(\epsilon^4).
\end{equation}
The rationale is as follows: while the rain amount in the DL remains comparable to that higher up in the troposphere, its \emph{change}, e.g., by evaporation below cloud base, occurring over an asymptotically shorter vertical distance, is sufficient to contribute to the stronger thermodynamical perturbations present in the DL. We emphasize that, even though we scaled the rain water mixing ratio as $O(\epsilon^4)$, this should \emph{not} be taken to be representative of synoptic-scale mean rainfall rates in the midlatitudes. Rather, it represents the rainfall rate that is \emph{directly due to synoptic-scale processes,} i.e., the synoptic precipitation regions arising from warm frontal lifting. According to data evaluated by \citet{austin1972}, shown in Table 5 of the cited work, this rate is actually quite small, in the range of $\sim0.5-1\text{mm/h}$, with embedded mesoscale precipitation areas contributing a much higher amount of rain. Thus, a model covering most realistic scenarios would have to account for multiple horizontal scales (see also \citep{HirtEtAl2023}). This is beyond the scope of the present work, however, and keeping only the resolved synoptic-scale processes, we restrict our scalings to \eqref{qr_scaled_pqg} and \eqref{qr_scaled_pdl}.
\subsubsection{Phase changes}
The adopted scalings of the mixing ratios in the PQG regime agree with \citet{baeumer2023}, except for the saturation value, $q_{\text{vs}}$. Accordingly, their scalings for the phase conversion terms will also carry over to our analysis, and here is their justification from a physical perspective:

The Kessler-style closures in \eqref{eq:GoverningEquationsSupplement-e}-\eqref{eq:GoverningEquationsSupplement-h} have been incorporated into models for moist convection before \citep{klein2006, hittmeir2018}. In in an analysis of moist \emph{large-scale} flows, the corresponding closures should therefore be interpreted as large-scale averages over the smaller scales at the appropriate order, although we do not construct a multiscale model here. In particular, using \eqref{eq:GoverningEquationsSupplement-e}-\eqref{eq:GoverningEquationsSupplement-h} in the PQG regime, the rate constants should \emph{not} be identified with those in the original closures. Rather, we consider their large-scale averages and, for simplicity, re-parameterize them in a form analogous to the original scheme. With this in mind, our central postulate is that \emph{all phase transitions are present in the asymptotic expansion at leading order} (in line with \citet{baeumer2023}).

As for the scaling in the DL, the lack of leading-order transport of cloud water due to \eqref{qc_scaled_dl} implies that the autoconversion term $S_{\text{ac}}$ should likewise drop out of the reduced model, as well as the collection term $S_{\text{cr}}$. Due to the assumption of stronger variability in relative humidity expressed in \eqref{qv_scaled_dl}, Eq.\ \eqref{eq:GoverningEquations-f} therefore reduces to the constraint that no condensation occurs at leading order. We will thoroughly explore the consequences of this systematic reduction in section~4.
\subsection{Closures for friction and turbulent mixing}
The impact of friction and turbulent mixing on large-scale atmospheric flow is certainly significant. In line with previous derivations of synoptic scale models, however, we will explicitly incorporate only the turbulent friction term $\mathcal{D}_{\vect{u}}$ in its textbook form in the Ekman layer. Other terms of the form $\mathcal{D}_\star$ are included in the derivation without discussing their concrete functional forms. Proceeding purely pragmatically, we assume that they enter the asymptotic expansions at the same order as the corresponding transport terms of the form $D_t(\cdot)$.
\subsection{Scaled governing equations}
Nondimensionalizing the governing equations and implementing the distinguished limits laid out above, we arrive at
\begin{subequations}\label{eq:GoverningEquationsDimless}
\begin{align}
\label{eq:GoverningEquationsDimless-a}
D_t\vect{u}+\frac{1}{\epsilon}f\vect{k}\times\vect{u}+\frac{1}{\epsilon^3}\frac{1}{\rho}\grad_\parallel{p}
   = & \epsilon^3\frac{\rho_d}{\rho}q_r V_r\partial_z\vect{u}+\mathcal{D}_{\vect{u}}, 
    \\
\label{eq:GoverningEquationsDimless-b}
D_tw+\frac{1}{\epsilon^5}\frac{1}{\rho}\partial_zp
  = & -\frac{1}{\epsilon^5} \nonumber \\
  & +\epsilon^3\frac{\rho_d}{\rho}q_r V_r\partial_z w+\mathcal{D}_w,
    \\
\label{eq:GoverningEquationsDimless-c}
\partial_t\rho_d+\nabla_\parallel\cdot(\rho_d\vect{u})+\partial_z(\rho_dw) 
  = & 0, 
    \\
\label{eq:GoverningEquationsDimless-d}
C_\epsilon D_t\ln\theta +\epsilon^3\Sigma_\epsilon D_t\ln p & \nonumber \\
+\epsilon\frac{L\phi_\epsilon}{T}D_tq_v
  = & \epsilon^2 k_lV_Tq_r \nonumber \\
  & \times(\partial_z\ln\theta+\epsilon\Gamma\partial_z\ln p) \nonumber \\
  & +Q+\mathcal{D}_\theta,
  \\
\label{eq:GoverningEquationsDimless-e}
D_tq_v
  = & S_{\text{ev}}-S_{\text{cd}}+\mathcal{D}_v, 
    \\
\label{eq:GoverningEquationsDimless-f}
D_tq_c
  = & S_{\text{cd}}-S_{\text{cr}}-S_{\text{ac}}+\mathcal{D}_c, 
    \\
\label{eq:GoverningEquationsDimless-g}
D_tq_r-\frac{1}{\epsilon}\frac{1}{\rho_d}\partial_z(\rho_dV_rq_r)
  = & \frac{1}{\epsilon}\left[S_{\text{cr}}+S_{\text{ac}}-S_{\text{ev}}\right]+\mathcal{D}_r,
\end{align}
\end{subequations}
where
\begin{subequations}\label{eq:GoverningEquationsSupplementDimless}
\begin{align}
\label{eq:GoverningEquationsSupplementDimless-a}
\rho
  & = \rho_d(1+\epsilon^2(q_v+\epsilon q_c+\epsilon^2q_r)), 
    \\
\label{eq:GoverningEquationsSupplementDimless-b}
p
  & = \rho_dT\left(1+\epsilon^2\frac{q_v}{E}\right), 
    \\
\label{eq:GoverningEquationsSupplementDimless-c}
T
  & = \theta p^{\epsilon\Gamma}\equiv\theta\pi, 
    \\
\label{eq:GoverningEquationsSupplementDimless-d}
C_\epsilon
  & = 1 + \epsilon^2 k_v q_v + \epsilon^2 (k_l(q_c + \epsilon q_r)), 
    \\
    \label{eq:GoverningEquationsSupplementDimless-e}
S_{\text{ev}}
  & = \bar{C}_{\text{ev}}\frac{p}{\rho}(q_{\text{vs}}-q_v)^+q_r, 
    \\
\label{eq:GoverningEquationsSupplementDimless-f}
S_{\text{cd}}
  & = \bar{C}_{\text{cn}}(q_v-q_{\text{vs}})^+q_{\text{cn}}+\bar{C}_{\text{cd}}(q_v-q_{\text{vs}})q_c, 
    \\
\label{eq:GoverningEquationsSupplementDimless-g}
S_{\text{ac}}
  & = \bar{C}_{\text{ac}}(q_c-q_{\text{ac}})^+, 
    \\
\label{eq:GoverningEquationsSupplementDimless-h}
S_{\text{cr}}
  & = \bar{C}_{\text{cr}}q_cq_r, 
    \\
\label{eq:GoverningEquationsSupplementDimless-i}
\Sigma_\epsilon 
  & = \epsilon \kappa_v q_v + \Gamma k_l(q_c+\epsilon q_r), 
    \\
\label{eq:GoverningEquationsSupplementDimless-j}
\phi_\epsilon
  & = 1-\frac{k_l-\epsilon k_v}{L}(T-1).
\end{align}
\end{subequations}
The scaled CC relation reads
\begin{equation}\label{CC_diff_qvs_formula_scaled}
    \frac{d \ln e_s}{dT}=\frac{1}{\epsilon}\frac{AL\phi_\epsilon(T)}{T^2}
    \qquad , \qquad
    q_{\text{vs}}=E\frac{e_s}{p-\epsilon^2 e_s}.
\end{equation}
Here, the bars in \eqref{eq:GoverningEquationsSupplementDimless-e}-\eqref{eq:GoverningEquationsSupplementDimless-h} indicate that the constants $\bar{C}_\star$ are not necessarily the same as those in the original system; the prefactor $1/\epsilon$ in \eqref{eq:GoverningEquationsDimless-g} raises the corresponding leading-order terms in the equation for $q_r$ to the same order in the asymptotic scheme as those in \eqref{eq:GoverningEquationsDimless-e} and \eqref{eq:GoverningEquationsDimless-f}.
\section{Asymptotic expansion and derivation of PQG-DL-Ekman}
We now proceed with the derivation of our model equations from a standard formal expansion ansatz, starting with the $\text{PQG}_{\text{DL}}$ regime for the bulk flow. At the end of each respective subsection, we recast the final result in dimensional form for the reader's convenience. In doing so, we adopt the following notational conventions: for any model variable $f$ that admits a decomposition into a leading-order background profile and a higher-order perturbation, we write $\bar{f}=\bar{f}(z)$ for the former and $\tilde{f}=\tilde{f}(t,\vect{x},z)$ for the latter. Perturbations that appear both in the quasigeostrophic regime, henceforth denoted by $\text{PQG}_{\text{DL}}$, and in the diabatic layer are assigned superscripts accordingly. We denote the material derivative with respect to the leading-order geostrophic velocity by
\begin{equation}
    D_t^g = \partial_t + \vect{u}\cdot\grad_\parallel,
\end{equation}
where it will always be clear from the context whether $\vect{u}$ stands for the horizontal velocity field in $\text{PQG}_{\text{DL}}$ or that in the DL. Please recall that the present regime is an \emph{update} of the $\text{PQG}_{\text{DL}}$ model of \citet{baeumer2023}, with significant differences in the transport of water vapor and its potential vorticity (PV) formulation. For that reason, we provide all important steps of the derivation. Also note that the assumptions inherent in $\text{PQG}_{\text{DL}}$ only hold at tropospheric heights $\gtrsim 3\text{ km}$, due to the presence of the intermediate diabatic layer.
\subsection{The $\text{PQG}_{\text{DL}}$ equations}
As usual, we begin by expanding all model variables in terms of the asymptotically small parameter $\epsilon$. In parallel to classical QG, we assume purely vertical background profiles for the thermodynamical variables $p$, $\rho$, $T$ and $\theta$, respectively, indicating the corresponding terms in the expansion by \emph{single} subscripts. If some background distribution is constant altogether at some given order, this is indicated by a \emph{double} subscript. Finally, perturbation variables with unrestricted dependence on $(t,\vect{x},z)$ carry bracketed superscripts.

For the model variables that also occur in dry air, we thus make the ansatz
\begin{subequations}\label{PQG_exp_dry}
\begin{align}
p
  & = p_0 + \epsilon p_1 + \epsilon^2 p^{(2)} + \epsilon^3 p^{(3)} + o(\epsilon^3), 
    \\
\pi 
  & = \pi_0 + \epsilon \pi_1 + \epsilon^2 \pi_2 + \epsilon^3 \pi^{(3)} + o(\epsilon^3), 
    \\
\rho
  & = \rho_0 + \epsilon\rho_1 + \epsilon^2\rho^{(2)} + o(\epsilon^2), 
    \\
\theta
  & = \theta_{00} + \epsilon\theta_1 + \epsilon^2\theta^{(2)} + o(\epsilon^2), 
    \\
\label{PQG_exp_dry_T}
T
  & = T_{00} + \epsilon T_1 + \epsilon^2 T^{(2)} + o(\epsilon^2), 
    \\
\vect{u}
  & = \vect{u}^{(0)} + \epsilon\vect{u}^{(1)} + o(\epsilon),
    \\
w 
  & = w^{(0)} + \epsilon w^{(1)} + o(\epsilon),
\end{align}
\end{subequations}
where the constants $T_{00}$, $\theta_{00}$ are not introduced ad hoc, but rather a consequence of the Newtonian limit \citep{klein2006,hittmeir2018}. The latter also implies that spatiotemporal perturbations in the expansion of the Exner pressure $\pi$ only start to come into play at third order, even though dynamical pressure perturbations are $O(\epsilon^2)$.

As for the moist constituents, we recall \eqref{q_vs_scaled}, \eqref{qv_background} and \eqref{qv_scaled_qg}, which translate to the ansatz
\begin{equation}\label{PQG_exp_qv}
    q_v=q_{\text{vs}0}+\epsilon q_v^{(1)}+o(\epsilon).
\end{equation}
Here, $q_{\text{vs}0}$ is the leading-order term in the asymptotic expansion of the saturation mixing ratio. This expansion results from the corresponding ansatz \eqref{PQG_exp_dry_T} for temperature, $T$, via the CC relation \eqref{CC_diff_qvs_formula_scaled}. Thus, the leading-order profile is readily obtained by inserting the former into the latter and keeping only $O(1)$ contributions:
\begin{equation}
    q_{\text{vs}0}(z)=E \frac{e_{s0}(z)}{p_0(z)} = E e_{s_{ref}} \frac{\exp{LA T_1(z)}}{p_0(z)}.
\end{equation}
\emph{Saturation} in this ansatz thus is equivalent to
\begin{equation}
    q_v^{(1)}>q_{\text{vs}}^{(1)},
\end{equation}
with the first-order correction $q_{\text{vs}}^{(1)}$ to the leading-order saturation mixing ratio. This correction depends on the temperature perturbation $T^{(2)}$ and thus is \emph{not} a given quantity, nor part of the static background. We will show the exact form of this dependence when deriving the reduced moisture balances later in this section. As already mentioned, the presence of the moist background profile $q_{\text{vs}0}$ restricts the applicability of our model to atmospheric regions with a uniformly large supply of water vapor.

\noindent
\emph{Remark:} It should be possible to overcome this restriction by introducing a \emph{transition} or \emph{internal layer} (in the sense of singular perturbation theory) at the interface separating dry and moist regions, with dry regions obeying classical QG. This is an endeavor that we plan to pursue in future work.

The liquid water species $q_c$ and $q_r$ straightforwardly expand as
\begin{subequations}\label{PQG_exp_liquid}
    \begin{align}
        q_c &= q_c^{(0)}+o(1), \\
        q_r &= q_r^{(0)}+o(1).
    \end{align}
\end{subequations}

Now, inserting the ansatz \eqref{PQG_exp_dry}-\eqref{PQG_exp_liquid} into the scaled governing equations and collecting terms of equal order in $\epsilon$, we can deduce:
\paragraph{Horizontal momentum and mass balances:}
Since friction does not \emph{directly} impact the bulk flow, i.e.~$\mathcal{D}_{\vect{u}}\equiv0$ outside the Ekman layer, the leading-order contributions to eqs.~\eqref{eq:GoverningEquationsDimless-a} and \eqref{eq:GoverningEquationsDimless-c} are formally equivalent to those in the model for dry air, except for the pressure perturbation $p^{(2)}$, which depends on the moist background distribution via \eqref{eq:GoverningEquationsSupplementDimless-b}. We therefore obtain geostrophic balance,
\begin{equation}\label{PQG_geostrophy_original}
    f_0\vect{k}\times\vect{u}^{(0)}=-\frac{1}{\rho_0}\grad_\parallel{p}^{(2)},
\end{equation}
from the horizontal momentum Eq.\ \eqref{eq:GoverningEquationsDimless-a} and incompressibility of the horizontal geostrophic velocity,
\begin{equation}
    \nabla_\parallel\cdot\vect{u}^{(0)}=0,
\end{equation}
as an immediate consequence. By a standard argument, evaluation of the continuity equation at leading order then yields the constraint
\begin{equation}\label{PQG_velocity_vert}
    w^{(0)}\equiv0,
\end{equation}
also characteristic of QG theory. At next order, we get
\begin{equation}
    \nabla_\parallel\cdot\vect{u}^{(1)}+\frac{1}{\rho_0}\partial_z(\rho_0w^{(1)})=0,
\end{equation}
which is an \emph{anelastic constraint.}

In the horizontal momentum equation, the $O(\epsilon)$ contributions read
\begin{align}
    D_t^{(0)}\vect{u}^{(0)} & +\beta y\vect{k}\times\vect{u}^{(0)}+f_0\vect{k}\times\vect{u}^{(1)} \nonumber \\
    & +\frac{1}{\rho_0}\grad_\parallel{p^{(3)}}-\frac{\rho_1}{\rho_0^2}\grad_\parallel{p^{(2)}}=0.
\end{align}
Taking the vertical component of the curl, this leads to 
\begin{align}
   & D_t^{(0)}\left[\zeta^{(0)}+\beta y\right]+f_0\nabla_\parallel\cdot\vect{u}^{(1)} \nonumber \\
   & =D_t^{(0)}\left[\zeta^{(0)}+\beta y\right]-\frac{f_0}{\rho_0}\partial_z(\rho_0w^{(1)})=0,
\end{align}
the transport equation for the absolute quasigeostrophic vorticity $\zeta^{(0)}+\beta y$, with the relative vorticity $\zeta^{(0)}:=\partial_x v^{(0)}-\partial_y u^{(0)}$. Here, $D_t^{(0)}=\partial_t+(\vect{u}^{(0)}\cdot\grad_\parallel)$ is the material derivative with respect to the geostrophic flow.
\paragraph{Vertical momentum balance:}
The vertical momentum Eq.~\eqref{eq:GoverningEquationsDimless-b} is dominated by hydrostatic balance at all relevant orders. Here, asymptotic expansion at the relevant orders and utilization of the ideal gas law \eqref{eq:GoverningEquationsSupplement-b} result in the leading-order explicit solution
\begin{equation}\label{PQG_hydrostat}
T_{00}\rho_0(z)=p_0(z)= p_0(0) e^{-\frac{z}{T_{00}}}
\end{equation}
and the relation
\begin{equation}
    \partial_z\left(\frac{p_1}{\rho_0}\right)=T_1=\theta_1-\pi_1
\end{equation}
for the static background. In \eqref{PQG_hydrostat}, we included the constant factors $T_{00} = p_0(0) = 1$ for clarity. At next order, we obtain
\begin{equation}
\partial_z\left(\frac{p^{(2)}}{\rho_0}\right)=\theta^{(2)}+\pi_1\theta_1+\pi_2+\frac{\rho_1}{\rho_0}T_1+\left(\frac{1}{E}-1\right)q_{\text{vs}0}
\end{equation}
as an expression for the leading-order \emph{buoyancy} perturbation. In order to get to the usual form of this relation, we again proceed as \citet{baeumer2023}, recognizing that all terms on the right-hand side except for $\theta^{(2)}$ depend on $z$ only - therefore, setting
\begin{equation}
    \phi^{(2)}:=\frac{p^{(2)}}{\rho_0}-\int_{0}^{z}\left[\pi_1\theta_1+\pi_2+\frac{\rho_1}{\rho_0}T_1+\left(\frac{1}{E}-1\right)q_{\text{vs}0}\right]d\zeta,
\end{equation}
we can write
\begin{equation}\label{PQG_buoyancy}
    \partial_z \phi^{(2)} = \theta^{(2)}
\end{equation}
as our buoyancy relation for the adjusted pressure perturbation $\tilde{\phi}$, scaled by the background density. We can also rewrite the geostrophic balance relation \eqref{PQG_geostrophy_original} as
\begin{equation}\label{PQG_geostrophy_final}
    f_0\vect{k}\times\vect{u}^{(0)}=-\grad_\parallel{\phi^{(2)}},
\end{equation}
since the horizontal gradient annihilates the added background terms.
\paragraph{Transport of potential temperature:}
Recalling \eqref{PQG_velocity_vert} and \eqref{PQG_hydrostat}, expanding \eqref{eq:GoverningEquationsDimless-d} yields
\begin{align}
    D_t^{(0)}\theta^{(2)} & +w^{(1)}\frac{d\theta_1}{dz}+LD_t^{(0)}q_v^{(1)} \nonumber \\
    & +Lw^{(1)}\frac{dq_{\text{vs}0}}{dz}=Q^{(0)}+\mathcal{D}_\theta^{(0)}.
\end{align}
Introducing the (linearized) equivalent potential temperature
\begin{equation}
    \theta_e:=\theta+Lq_v,
\end{equation}
we can equivalently write
\begin{equation}\label{PQG_theta_equiv}
    D_t^{(0)}\theta_e^{(2)}+w^{(1)}\frac{d\theta_{e1}}{dz}=Q^{(0)}+\mathcal{D}_\theta^{(0)}.
\end{equation}
We remark that, while we do not consider the inclusion of turbulent mixing crucial for the model, it might be necessary for its rigorous mathematical validation, since the thermodynamics encoded in \eqref{PQG_theta_equiv} depend on the nonsmooth phase changes in the reduced version of \eqref{eq:GoverningEquationsDimless-e}.
\paragraph{Transport of water vapor:}
With the scaling assumptions laid out in section~3, we obtain
\begin{align}
    D_t^{(0)}q_v^{(1)}+w^{(1)}\frac{dq_{\text{vs}0}}{dz}=&\bar{C}_{\text{ev}}T_{00}\left[q_{\text{vs}}^{(1)}-q_v^{(1)}\right]^+q_r^{(0)} \nonumber \\
    &-\bar{C}_{\text{cn}}\left[q_v^{(1)}-q_{\text{vs}}^{(1)}\right]^+q_{\text{cn}}^{(0)} \nonumber \\
    &-\bar{C}_{\text{cd}}\left[q_v^{(1)}-q_{\text{vs}}^{(1)}\right]q_c^{(0)}+\mathcal{D}_v^{(0)},
\end{align}
where the turbulent closure $\mathcal{D}_v^{(0)}$ again might prove helpful in order to prove the general existence of (strong) solutions, as in previous work on moist transport equations with phase changes \citep{hittmeir2017,hittmeir2020,hittmeir2023,doppler2024}. Remember that $q_v$ and $q_{\text{vs}}$ are identical at leading order, which is why the first-order correction $q_{\text{vs}}^{(1)}$ to the saturation mixing ratio shows up on the right-hand side. This quantity can be written in the form
\begin{equation}
    E \frac{e_s^{(1)}}{p_0}-E\frac{p_1}{p_0^2}e_{s0},
\end{equation}
where
\begin{equation}
    e_s^{(1)}=A \left[L T^{(2)}-\left(L+\frac{k_l+k_v}{2}\right)T_1^2\right]e_{s0}.
\end{equation}
Thus, $q_{\text{vs}}^{(1)}$ varies linearly with the unknown $T^{(2)}$ or, equivalently, $\theta^{(2)}$, and the saturation threshold is \emph{unknown} a priori.
\paragraph{Transport of cloud water:}
Since there is no ``background liquid water'', the leading-order transport of cloud water is purely horizontal:
\begin{align}
    D_t^{(0)}q_c^{(0)}=&\bar{C}_{\text{cn}}\left[q_v^{(1)}-q_{\text{vs}}^{(1)}\right]^+q_{\text{cn}}^{(0)}+\bar{C}_{\text{cd}}\left[q_v^{(1)}-q_{\text{vs}}^{(1)}\right]q_c^{(0)} \nonumber \\
    &-\bar{C}_{\text{ac}}\left[q_c^{(0)}-q_{\text{ac}}^{(0)}\right]^+ -\bar{C}_{\text{cr}}q_c^{(0)}q_r^{(0)} + \mathcal{D}_c^{(0)},
\end{align}
where the same remarks as above apply with regard to the sources on the right-hand side.
\paragraph{Quasi-steady column of rain:}
From the scaled Eq.\ \eqref{eq:GoverningEquationsDimless-g}, we obtain - as in section~6 of \citet{baeumer2023} - a \emph{diagnostic} relationship for the mixing ratio of rain at leading order:
\begin{align}\label{PQG_qr}
    -\frac{1}{\rho_0}\partial_z(\rho_0 V_T q_r^{(0)})=&\bar{C}_{\text{ac}}\left[q_c^{(0)}-q_{\text{ac}}^{(0)}\right]^+ + \bar{C}_{\text{cr}}q_c^{(0)}q_r^{(0)} \nonumber \\
    &-\bar{C}_{\text{ev}}T_{00}\left[q_{\text{vs}}^{(1)}-q_v^{(1)}\right]^+q_r^{(0)}.
\end{align}
This equation exhibits a mathematically straightforward, but in this context unusual structure: since it is essentially a first-order ODE that depends on $(t,\vect{x})$ only parametrically, the only data that we need to prescribe in order to specify a unique solution is
\[
    q_r^{(0)}(t,\vect{x},h)
\]
at a fixed height $h$, for all $(t,\vect{x})$. The physically sensible choice here is to set the rain water mixing ratio to zero at the top of the layer, since (almost) all precipitation in the earth's atmosphere is generated below the tropopause.

We now present the $\text{PQG}_{\text{DL}}$ system in its preliminary form:
\begin{subequations}\label{PQG_preliminary}
    \begin{align}
        f_0\vect{k}\times\vect{u}^{(0)} &= -\grad_\parallel{\phi^{(2)}}, \\
        \partial_z \phi^{(2)} &= \theta^{(2)}, \\
        D_t^{(0)}\left[\zeta^{(0)}+\beta y\right]&=\frac{f_0}{\rho_0}\partial_z(\rho_0w_1), \\
        \label{PQG_preliminary_theta_e}
        D_t^{(0)}\theta_e^{(2)}+w^{(1)}\frac{d\theta_{e1}}{dz}&=Q^{(0)}+\mathcal{D}_\theta^{(0)}, \\
        \label{PQG_preliminary_qv}
        D_t^{(0)}q_v^{(1)}+w^{(1)}\frac{dq_{\text{vs}0}}{dz}&=S_{\text{ev}}^{(0)}-S_{\text{cd}}^{(0)}+\mathcal{D}_v^{(0)}, \\
        \label{PQG_preliminary_qc}
        D_t^{(0)}q_c^{(0)}&=S_{\text{cd}}^{(0)}-S_{\text{ac}}^{(0)}-S_{\text{cr}}^{(0)}+\mathcal{D}_c^{(0)}, \\
        \label{PQG_preliminary_qr}
        -\frac{1}{\rho_0}\partial_z(\rho_0 V_T q_r^{(0)})&=S_{\text{ac}}^{(0)}+S_{\text{cr}}^{(0)}-S_{\text{ev}}^{(0)},
    \end{align}
\end{subequations}
with
\begin{subequations}
    \begin{align}
        q_{\text{vs}}^{(1)} &= \left[AL(\theta^{(2)}+\pi_1\theta_1+\pi_2)-A(L+k_l/2)T_1^2-\frac{p_1}{p_0}\right] q_{\text{vs}0} \label{qvs_pert} \\
        S_{\text{ev}}^{(0)}&=\bar{C}_{\text{ev}}T_{00}\left[q_{\text{vs}}^{(1)}-q_v^{(1)}\right]^+q_r^{(0)} \label{PQG_DL_qvs_pert} \\
        S_{\text{cd}}^{(0)}&=\bar{C}_{\text{cn}}\left[q_v^{(1)}-q_{\text{vs}}^{(1)}\right]^+q_{\text{cn}}^{(0)}+\bar{C}_{\text{cd}}\left[q_v^{(1)}-q_{\text{vs}}^{(1)}\right]q_c^{(0)} \\
        S_{\text{ac}}^{(0)}&=\bar{C}_{\text{ac}}\left[q_c^{(0)}-q_{\text{ac}}^{(0)}\right]^+ \\
        S_{\text{cr}}^{(0)}&=\bar{C}_{\text{cr}}q_c^{(0)}q_r^{(0)}.
    \end{align}
\end{subequations}
Here, all (subscripted) background terms are given, and the turbulent closures $\mathcal{D}_\star$ can be specified, e.g., as proportional to (horizontal) Laplacians of the respective transported quantities when needed.

Once more analogous to classical QG (and previous investigations of PQG dynamics as in, e.g., \citep{wetzel2019}), we aim to reformulate this system based on a suitable notion of \emph{potential vorticity} (PV). This task will be a fair bit more involved than in the dry regime, owing to the presence of the various moisture species and phase changes.
\subsubsection*{Potential vorticity formulation of $\text{PQG}_{\text{DL}}$}
We recall the \emph{quasigeostrophic potential vorticity,} here indicated by PV for simplicity, and given by
\begin{equation}
    \text{PV}=\zeta^{(0)}+\beta y+\frac{f_0}{\rho_0}\partial_z\left(\frac{\rho_0\theta^{(2)}}{d\theta_1/dz}\right).
\end{equation}
It permits a formulation of classical QG in terms of only one horizontal transport equation,
\begin{equation}
    D_t^{(0)} \text{PV} =\frac{f_0}{\rho_0}\partial_z\left(\frac{\rho_0Q^{(0)}}{d\theta_1/dz}\right),
\end{equation}
while the pressure perturbation $\phi^{(2)}$ (and consequently $\vect{u}^{(0)}$ and $\theta^{(2)}$) can be recovered from the linear Poisson-type equation
\begin{equation}\label{qg_recovery}
    \frac{1}{f_0}\Delta_\parallel\phi^{(2)}+\frac{f_0}{\rho_0}\partial_z\left(\frac{\rho_0\partial_z\phi^{(2)}}{d\theta_1/dz}\right)= \text{PV}-\beta y,
\end{equation}
given appropriate boundary conditions. In particular, this reformulation completely eliminates the small QG vertical velocity $w^{(1)}$ from the system, which actively shapes the dynamics solely via the bottom boundary condition. This becomes relevant when we couple QG to a boundary layer, that is, the Ekman layer - we will thoroughly explain the analogous situation in PQG-DL-Ekman in the next two subsections.

Now, we intend to find an analogous formulation for the $\text{PQG}_{\text{DL}}$ system \eqref{PQG_preliminary}. Proceeding along similar lines as \citet{smith2017}, we need to determine a suitable notion of PV analogous to the QG potential vorticity. Here, basing PV on the equivalent potential temperature $\theta_e$ seems natural, since its evolution equation \eqref{PQG_preliminary_theta_e} mirrors that of $\theta$ in dry air. We therefore define $\text{PV}_e$ as
\begin{equation}
    \text{PV}_e=\zeta^{(0)}+\beta y+\frac{f_0}{\rho_0}\partial_z\left(\frac{\rho_0\theta_e^{(2)}}{d\theta_{e1}/dz}\right).
\end{equation}
Observing that all equations involved in the derivation of the evolution equation of $\text{PV}_e$ are (up to source terms) the same in our model as in the original PQG equations \citep{smith2017}, we immediately arrive at
\begin{align}
    D_t^{(0)}\text{PV}_e= & -\frac{f_0}{d\theta_{e1}/dz}\partial_z\vect{u}^{(0)}\cdot\grad_\parallel\theta_e^{(2)} \nonumber \\
    & +\frac{f_0}{\rho_0}\partial_z\left(\frac{\rho_0(Q^{(0)}+\mathcal{D}_\theta^{(0)})}{d\theta_{e1}/dz}\right).
\end{align}
The presence of $d\theta_{e1}/dz$ in the denominator indicates that the validity of this formulation requires an \emph{unconditionally stably stratified} atmosphere in the large-scale mean, i.e.
\begin{equation}\label{PQG_stability}
    \frac{d\theta_{e1}}{dz}>0
\end{equation}
throughout. This condition might seem quite restrictive - however, thanks to the inclusion of the diabatic layer, we only need it to apply at altitudes greater than $\sim3\text{ km}$, while even neutral stratification can be attained below \citep{klein2022}.

\noindent
\emph{Remark:} If one wants to preserve the less restrictive condition $d\theta_1/dz>0$, one can still revert to ``dry'' PV as a dynamical variable and rewrite the $\text{PQG}_{\text{DL}}$ equations accordingly. We should mention that, as pointed out to the authors by Sam Stechmann, \eqref{PQG_stability} should still be assumed in any formulation of the model in order to guarantee the validity of QG theory in the first place.

Clearly, due to the incorporation of the moist constituents $q_v$, $q_c$ and $q_r$, we need more than one transport equation to fully determine the $\text{PQG}_{\text{DL}}$ flow. We would like to not use $q_v^{(1)}$, whose transport equation \eqref{PQG_preliminary_qv} depends on $w^{(1)}$, and therefore, again following \citet{smith2017}, introduce an auxiliary moist variable: defining
\begin{equation}\label{PQG_M}
    \tilde{M}:=\theta_e^{(2)}-\frac{d\theta_{e1}/dz}{d q_{\text{vs}0}/dz}q_v^{(1)}\equiv\theta_e^{(2)}+B(z)q_v^{(1)},
\end{equation}
it is straightforward to check that
\begin{equation} \label{PQG_DL_M}
    D_t^{(0)}\tilde{M}=B(z)\left[S_{\text{ev}}^{(0)}-S_{\text{cd}}^{(0)} + \mathcal{D}_v^{(0)}\right] + \mathcal{D}_\theta^{(0)} + Q^{(0)},
\end{equation}
which does not involve the vertical velocity.

The evolution equation for cloud water \eqref{PQG_preliminary_qc} and the diagnostic relation \eqref{PQG_preliminary_qr} for rain do not depend on $w^{(1)}$ in the first place, so no reformulation is needed there.

Next, we turn to the moist analogue of the elliptic recovery equation \eqref{qg_recovery} for the pressure perturbation $\phi^{(2)}$. Here, a linear combination of $\text{PV}_e$ and the vertical derivative of $\tilde{M}$ does the trick: by direct calculation, one can verify that
\begin{align} \label{PQG_DL_inversion}
    \frac{1}{f_0}\Delta_\parallel\phi^{(2)}+\frac{f_0}{\rho_0}\partial_z\left(\frac{\rho_0}{d\theta_{e1}/dz}\frac{B(z)}{L+B(z)}\partial_z\phi^{(2)}\right) \nonumber \\
    =\text{PV}_e-\frac{f_0}{\rho_0}\partial_z\left(\frac{\rho_0}{d\theta_{e1}/dz}\frac{L}{L+B(z)}\tilde{M}\right)-\beta y,
\end{align}
which, given $\text{PV}_e$ and $\tilde{M}$, is a linear elliptic equation for $\phi^{(2)}$. (The term $B(z)$, implicitly defined in \eqref{PQG_M}, will always be positive when \eqref{PQG_stability} holds.) We should clarify that linearity here holds for given \emph{balanced} or low-frequency components of the flow, which in this context means that their respective evolution equations do not incorporate vertical advection; see \citet{wetzel2019} for a thorough discussion of this concept. The PV formulations of \citet{smith2017} and \citet{wetzel2019} are \emph{nonlinear} when only balanced quantities are given. This nonlinearity is, in fact, a consequence of the fast phase transitions that the authors prescribe from the outset. The present full microphysics approach - while more complicated overall - therefore leads to a structurally \emph{simpler} inversion equation for the stream function, that is, the pressure perturbation. In particular, Eq.~\eqref{PQG_DL_inversion} does not exhibit a jump in the (background) buoyancy frequency, which physically is caused by the passage of air parcels from undersaturated to saturated air. This might be puzzling at first glance, since $d\theta/dz$ and $d\theta_e/dz$ \emph{do} differ at $O(\epsilon)$ also in the present scaling. In the context of QG theory with time resolved, non-singular source terms, however, \emph{vertical motion is too weak to make this jump affect the temperature of an idealized rising or falling air parcel at the same order.} This discontinuity can therefore be expected to come into play only at the level of the dynamical perturbations, which can be seen from the structure of \eqref{PQG_DL_M} and \eqref{PQG_DL_inversion} ($\partial_z \tilde{M}$ will be discontinuous at a horizontally aligned phase interface).

The transition from our full microphysics model to that of Wetzel et al.\ can formally be achieved by altering our asymptotic scheme to include a prefactor of the form $1/\epsilon^n$ with sufficiently large exponent, $n$, in the condensation term; in order to clarify the connection between the two approaches, we sketch the corresponding derivation in the Appendix.

In conclusion, recasting all equations in dimensional form as described at the beginning of this section, we obtain
\begin{subequations} \label{PQG_DL_dim}
    \begin{align}
        D_t^g \text{PV}_e =& -\frac{f}{d\bar{\theta}_e/dz} \partial_z \vect{u}^{\text{QG}} \cdot \grad_\parallel \left(\frac{L_{\text{ref}}}{c_{\text{pd}}}\tilde{q}_v^{\text{QG}}\right) \nonumber \\
        &+ \frac{f}{\bar{\rho}} \partial_z \left(\frac{\bar{\rho}\left(Q^{\text{QG}}+\mathcal{D}_\theta^{\text{QG}}\right)}{d\bar{\theta}_e/dz}\right) \\
        D_t^g \tilde{M} =& B(z) \left[S_{\text{ev}}^{\text{QG}} - S_{\text{cd}}^{\text{QG}} + \mathcal{D}_v^{\text{QG}}\right] + Q^{\text{QG}} + \mathcal{D}_\theta^{\text{QG}} \\
        D_t^g q_c^{\text{QG}} =& S_{\text{cd}}^{\text{QG}} - S_{\text{ac}}^{\text{QG}} - S_{\text{cr}}^{\text{QG}} + \mathcal{D}_c^{\text{QG}} \\
        -\frac{1}{\bar{\rho}} \partial_z \left(\bar{\rho} V_r q_r^{\text{QG}}\right) =& S_{\text{ac}}^{\text{QG}} + S_{\text{cr}}^{\text{QG}} - S_{\text{ev}}^{\text{QG}} \\
        \frac{1}{f}\Delta_\parallel \tilde{\phi}^{\text{QG}} + \frac{f}{\bar{\rho}} \partial_z & \left(\frac{\bar{\rho}}{d\bar{\theta}_e/dz}\frac{B(z)}{L_{\text{ref}}/c_{\text{pd}}+B(z)} \frac{\theta_{\text{ref}}}{g} \partial_z \tilde{\phi}^{\text{QG}}\right) \nonumber \\
         = & \text{PV}_e - \beta y \nonumber \\
         &  - \frac{f}{\bar{\rho}} \partial_z \left(\frac{\bar{\rho}}{d\bar{\theta}_e/dz}\frac{L_{\text{ref}}/c_{\text{pd}}}{L_{\text{ref}}/c_{\text{pd}}+B(z)} \tilde{M}\right) \label{PQG_DL_dim_inversion} \\
        f \vect{k} \times \vect{u}^{\text{QG}} =& -\grad_\parallel \tilde{\phi}^{\text{QG}} \\
        \partial_z \tilde{\phi}^{\text{QG}} =& g\frac{\tilde{\theta}^{\text{QG}}}{\theta_{\text{ref}}},
    \end{align}
\end{subequations}
with the additional relations
\begin{subequations} \label{PQG_DL_dim_supplement}
    \begin{align}
        B(z) &= -\frac{d\bar{\theta}_e/dz}{d\bar{q}_{\text{vs}}/dz} \\
        \tilde{q}_v^{\text{QG}} &= \frac{1}{L_{\text{ref}}/c_{\text{pd}}+B(z)} \left(\tilde{M} - \tilde{\theta}^{\text{QG}}\right) \\
        \tilde{q}_{\text{vs}}^{\text{QG}} &= \frac{L_{\text{ref}}}{R_v T_{\text{ref}}} \frac{\tilde{\theta}^{\text{QG}}}{\theta_{\text{ref}}} \bar{q}_v + s^{\text{QG}}(z) \\
        S_{\text{ev}}^{\text{QG}} &= \bar{C}_{\text{ev}} T_{\text{ref}} \left(\tilde{q}_{\text{vs}}^{\text{QG}}-\tilde{q}_v^{\text{QG}}\right)^+ q_r^{\text{QG}} \\
        S_{\text{cd}}^{\text{QG}} &= \bar{C}_{\text{cn}} \left(\tilde{q}_v^{\text{QG}}-\tilde{q}_{\text{vs}}^{\text{QG}}\right)^+ q_{\text{cn}}^{\text{QG}} + \bar{C}_{\text{cd}} \left(\tilde{q}_v^{\text{QG}}-\tilde{q}_{\text{vs}}^{\text{QG}}\right) q_c^{\text{QG}} \\
        S_{\text{ac}}^{\text{QG}} &= \bar{C}_{\text{ac}} \left(q_c^{\text{QG}}-q_{\text{ac}}\right)^+ \\
        S_{\text{cr}}^{\text{QG}} &= \bar{C}_{\text{cr}} q_c^{\text{QG}} q_r^{\text{QG}}.
    \end{align}
\end{subequations}
Here, $s(z)$ represents all the higher-order background contributions to $\tilde{q}_{\text{vs}}^{\text{QG}}$, cf.~\eqref{qvs_pert}.
\subsection{Precipitating DL dynamics}
All (P)QG models rely on a stable background stratification of $\theta$ or $\theta_{e}$ and small deviations of these variables from their background states of, at most, order $O(\epsilon^2)$. As a consequence, the diabatic heating rates they permit are quite modest, in the range of $\sim 3\;\text{K/day}$. As argued by \citet{klein2022}, such heating rates agree well with the synoptic-scale averages found in reanalysis data across most of the free troposphere - see also the references mentioned in the cited work. Near the surface, however, diabatic effects can be seen to significantly exceed the strength that QG scaling implicitly assumes. This observation motivated the introduction of the diabatic layer equations by Klein et al.~to extend the classical QG-Ekman model to a triple-deck boundary layer theory in which the layer of intermediate height is both dynamically and thermodynamically active. In the sequel, we show how the ``dry'' DL equations that incorporate strong diabatic heating only via an external heat source can be systematically augmented by moist (thermo)dynamics, asymptotically matched to the $\text{PQG}_{\text{DL}}$ flow just derived in an overlap region.

The characteristic height of the DL can be estimated as $\sim \sqrt{\epsilon} h_{\text{sc}}$, which corresponds to a physical value of about $3\;\text{km}$. Adopting the notation of \citet{klein2022}, we abbreviate $\delta = \sqrt{\epsilon}$ and introduce the stretched layer coordinate $\eta = z/\delta$. This suggests the generic ansatz
\begin{equation}
    f^{\text{DL}} = f^{\text{DL}}(t,\vect{x},\eta) = f^{(0/2)}(t,\vect{x},\eta) + \delta f^{(1/2)}(t,\vect{x},\eta) +\dots
\end{equation}
for all DL model variables, where superscripts are written in the form $(i/2)$, both to signal that the DL variables are expanded in powers of $\sqrt{\epsilon}$, and to distinguish them from the corresponding (P)QG variables. In the original DL equations of \citet{klein2022}, the diabatic layer was characterized by the assumption that potential temperature perturbations could \emph{dynamically change} the lower-level stratification of the troposphere, e.g., toward a neutral state. Mathematically, this implied an expansion of the (nondimensional) form
\begin{equation}
    \theta^{\text{DL}}=1+\delta^2\theta_1(0)+\delta^3\theta^{(3/2)}+o(\delta^3),
\end{equation}
with $\theta^{(3/2)}=\theta^{(3/2)}(t,\vect{x},\eta)$.

As we have seen in the derivation of $\text{PQG}_{\text{DL}}$, the equivalent potential temperature $\theta_e=\theta+L q_v$ plays a role in the moist dynamics that is largely analogous to that of $\theta$ in the dry regime. Therefore, it seems plausible that $\theta_e$ should expand in the same fashion in a moist DL regime, and this is compatible with the scaling assumption on $\tilde{q}_v^{\text{DL}}$ expressed in \eqref{qv_scaled_dl}. Since equivalent potential temperature is linked to the evolution of the moist constituents via latent heating, we will obtain reduced dynamics with a two-way coupling of moist processes and (thermo)dynamics. Thus, the equations derived below constitute a systematic extension of the DL system of \citet{klein2022}, in which moisture could be incorporated only as a heat source. For the reader primarily interested in the qualitative nature of the interactions with the bulk flow, Fig.~\ref{fig:feedbacks_visual} at the end of this section provides a visual illustration of said interactions.

We thus state the formal expansion ansatz for the precipitating DL (PDL) dynamics without further ado:
\begin{subequations}\label{PDL_exp}
    \begin{align}
        p^{\text{DL}}= & p_{0/2}+\delta p_{1/2}+\delta^2 p_{2/2}+\delta^3 p_{3/2}+\delta^4 p^{(4/2)} \nonumber \\
        & +\delta^5 p^{(5/2)}+o(\delta^5), \\
        \rho^{\text{DL}}= & \rho_{0/2}+\delta \rho_{1/2}+\delta^2 \rho_{2/2}+\delta^3 \rho^{(3/2)}+o(\delta^3), \\
        \theta^{\text{DL}}= & 1+\delta^2\theta_{2/2}+\delta^3\theta^{(3/2)}+o(\delta^3), \\
        T^{\text{DL}}= & 1+\delta^2 T_{2/2}+\delta^3 T^{(3/2)}+o(\delta^3), \\
        \vect{u}^{\text{DL}}= & \vect{u}^{(0/2)}+\delta\vect{u}^{(1/2)}+o(\delta), \\
        w^{\text{DL}}= & w^{(0/2)}+\delta w^{(1/2)}+\delta^2 w^{(2/2)}+o(\delta^2), \\
        q_v^{\text{DL}}= & q_{\text{vs}0/2}+\delta q_v^{(1/2)}+o(\delta), \\
        q_c^{\text{DL}}= & q_c^{(0/2)}+o(1), \\
        q_r^{\text{DL}}= & q_r^{(0/2)}+o(1).
    \end{align}
\end{subequations}
Here, we have omitted terms in the expansions of the temperature variables that can easily be shown to vanish by preliminary matching considerations. The ansatz \eqref{PDL_exp} can now be inserted into \eqref{eq:GoverningEquationsDimless}, where we only need to rewrite any vertical derivatives  in terms of $\eta$:
\paragraph{Horizontal momentum and mass balances:} As in the ``dry'' DL equations of \citet{klein2022}, geostrophic balance is preserved in the diabatic layer at all relevant orders. We obtain
\begin{equation}\label{PDL_geostrophy_1}
    f_0\vect{k}\times\vect{u}^{(0/2)}=-\grad_\parallel{p^{(4/2)}}
\end{equation}
and
\begin{equation}\label{PDL_geostrophy_2}
    f_0\vect{k}\times\vect{u}^{(1/2)}=-\grad_\parallel{p^{(5/2)}},
\end{equation}
by inserting the expansion ansatz \eqref{PDL_exp} into the horizontal momentum Eq.~\eqref{eq:GoverningEquationsDimless-a}.

Evaluation of the continuity equation at the relevant orders provides constraints on the vertical velocity: again following \citet{klein2022}, we get
\begin{equation}
    w^{(0/2)}=w^{(1/2)}=0
\end{equation}
and
\begin{equation}
    \partial_\eta w^{(2/2)}=0,
\end{equation}
i.e., a vertical velocity that is imprinted on (P)DL by the bulk flow and stays constant throughout the layer:
\begin{equation}
    w^{(2/2)}=w^{(2/2)}(t,\vect{x})=w^{(1)}(t,\vect{x},0).
\end{equation}
\paragraph{Vertical momentum balance:} Here, the expansion of the static background (up to $p_{3/2}$) is still identical to that of the dry DL regime, and of no particular interest. At the level of the perturbation, a closer look is warranted:
\begin{equation}
    \partial_\eta p^{(4/2)}=-\rho^{(3/2)}=\theta^{(3/2)}-p_{3/2}-(\Gamma+\theta_1(0))\,\eta
\end{equation}
again suggests the introduction of an adjusted pressure perturbation in order to avoid unnecessary clutter:
\begin{equation}\label{PDL_pressure_modified}
    \phi^{(4/2)} := p^{(4/2)}+\int_0^\eta \left(p_{3/2}+(\Gamma+\theta_1(0))\,\eta'\right) d\eta'.
\end{equation}
We can now write
\begin{subequations}
    \begin{align}
        f_0\vect{k}\times\vect{u}^{(0/2)}&=-\grad_\parallel{\phi^{(4/2)}}, \\
        \partial_\eta\phi^{(4/2)}&=\theta^{(3/2)}
    \end{align}
\end{subequations}
for the geostrophic and hydrostatic balances in PDL.
\paragraph{Transport of potential temperature:} Here, before we derive the reduced equation, written in terms of the equivalent potential temperature $\theta_e$, we need to make clear the expansion of the saturation mixing ratio in the diabatic layer: systematic expansion of \eqref{CC_diff_qvs_formula} and \eqref{es_exact} yields
\begin{equation}
    q_{\text{vs}0/2} = E e_{s_{00}} = E e^{LA\theta_1(0)}
\end{equation}
and
\begin{equation} \label{PDL_qvs_pert}
    q_{\text{vs}}^{(1/2)}=ELA e_{s_{00}} T^{(3/2)}+E e_{s_{00}} \eta;
\end{equation}
the static moist background is thus \emph{constant} in the precipitating DL, while the first-order correction that determines the local saturation threshold again depends on the temperature perturbation and consequently is part of the \emph{output} of the PDL system.

Due to the constancy of $q_{\text{vs}0/2}$, vertical advection drops out of the leading-order balance and the reduced version of \eqref{eq:GoverningEquationsDimless-d} now reads
\begin{equation}
    (\partial_t+\vect{u}^{(0/2)}\cdot\nabla_\parallel)\theta_e^{(3/2)}\equiv D_t^{(0/2)}\theta_e^{(3/2)}=Q^{(0/2)}+\mathcal{D}_\theta^{(0/2)},
\end{equation}
where any heat sources that one might consider beyond latent heating resp.\ cooling are now permitted to attain (dimensional) rates
\begin{equation}
    Q^{\text{DL}}\sim\frac{\delta^3 T_{\text{ref}}}{t_{\text{ref}}}\approx 10\text{ K/day}.
\end{equation}
\paragraph{Moist constituents:} In determining the evolution of the moisture species $q_v^{\text{DL}}$, $q_c^{\text{DL}}$ and $q_r^{\text{DL}}$, we first need to resolve one crucial issue: by our scaling assumptions, the evolution equation for cloud water reduces to the constraint
\begin{align}
    S_{\text{cd}}^{(0/2)}= & \bar{C}_{\text{cn}}\left[q_v^{(1/2)}-q_{\text{vs}}^{(1/2)}\right]^+q_{\text{cn}}^{(0/2)} \nonumber \\
    & +\bar{C}_{\text{cd}}\left[q_v^{(1/2)}-q_{\text{vs}}^{(1/2)}\right]q_c^{(0/2)}=0,
\end{align}
which leads to the alternative
\begin{align}\label{qc_constraint_dl}
\left\{\begin{array}{cc}
   q_v^{(1/2)}=q_{\text{vs}}^{(1/2)}  & \text{in saturated air} \\
   q_c^{(0/2)}=0  & \text{in undersaturated air.}
\end{array}\right.
\end{align}
Since we have not imposed any restrictions on the generation of leading-order cloud water in $\text{PQG}_{\text{DL}}$, this constraint will necessitate additional conditions on the initial data for water vapor perturbations in the DL to guarantee matching to the bulk flow. First, let us discuss the saturated case:

Since we require $q_v^{(1/2)}=q_{\text{vs}}^{(1/2)}$ in saturated regions, the initial data for water vapor (and potential temperature) need to be chosen such that
    \begin{equation}
        q_v^{(1/2)}\vert_{t=0} \leq q_{\text{vs}}^{(1/2)}\vert_{t=0}
    \end{equation}
everywhere. Due to the structure of the (independently derived) Eqs.\ \eqref{PDL_dim_temp} and \eqref{PDL_dim_qv} (see below), this constraint will then be obeyed for all times, as long as we choose the heat source $Q^{(0/2)}$ such that no (leading-order) external cooling is applied to saturated air parcels.
 
In the respective balances for $q_v$ and $q_r$, only the evaporation term survives at leading order. Additionally, the vertical velocity drops out of \eqref{eq:GoverningEquationsDimless-e}, since the saturation mixing ratio becomes constant at leading order in the DL, and the resulting moisture dynamics in PDL read
\begin{subequations}
    \begin{align}
        D_t^{(0/2)}q_v^{(1/2)}&=S_{\text{ev}}^{(0/2)}+\mathcal{D}_v^{(0/2)}, \\
        S_{\text{cd}}^{(0/2)}&=0, \\
        -\partial_\eta(V_T q_r^{(0/2)})&=-S_{\text{ev}}^{(0/2)}.
    \end{align}
\end{subequations}
We can now state the complete, closed set of equations in dimensional form,
\begin{subequations} \label{PDL_dim}
    \begin{align}
        D_t^g \tilde{\theta}_e^{\text{DL}} &= Q^{\text{DL}} + \mathcal{D}_\theta^{\text{DL}} \label{PDL_dim_temp} \\
        D_t^g \tilde{q}_v^{\text{DL}} &= S_{\text{ev}}^{\text{DL}} + \mathcal{D}_v^{\text{DL}} \label{PDL_dim_qv} \\
        0 &= S_{\text{cd}}^{\text{DL}} \label{PDL_dim_qc} \\
        -\partial_z(V_r q_r^{\text{DL}}) &= -S_{\text{ev}}^{\text{DL}} \label{PDL_dim_qr} \\
        f \vect{k} \times \vect{u}^{\text{DL}} &= -\grad_\parallel \tilde{\phi}^{\text{DL}} \\
        \partial_z \tilde{\phi}^{\text{DL}} &=g \frac{\tilde{\theta}^{\text{DL}}}{\theta_{\text{ref}}},
    \end{align}
\end{subequations}
with the supplementary relations
\begin{subequations}\label{PDL_dim_supplement}
    \begin{align}
    \tilde{\theta}^{\text{DL}} &= \tilde{\theta}_e^{\text{DL}} - \frac{L_{\text{ref}}}{c_{\text{pd}}}\tilde{q}_v^{\text{DL}}, \\
    \tilde{q}_{\text{vs}}^{\text{DL}} &= \frac{L_{\text{ref}}}{R_v T_{\text{ref}}} \frac{\tilde{\theta}^{\text{DL}}}{\theta_{\text{ref}}} q_{\text{vs}_{\text{ref}}} + s^{\text{DL}}(z), \\
    S_{\text{ev}}^{\text{DL}} &= \bar{C}_{\text{ev}} T_{\text{ref}} \left(\tilde{q}_{\text{vs}}^{\text{DL}}-\tilde{q}_v^{\text{DL}}\right)^+ q_r^{\text{DL}} \\
    S_{\text{cd}}^{\text{DL}} &= \bar{C}_{\text{cn}} \left(\tilde{q}_v^{\text{DL}}-\tilde{q}_{\text{vs}}^{\text{DL}}\right)^+ q_{\text{cn}}^{\text{DL}} + \bar{C}_{\text{cd}} \left(\tilde{q}_v^{\text{DL}}-\tilde{q}_{\text{vs}}^{\text{DL}}\right) q_c^{\text{DL}},
    \end{align}
\end{subequations}
where $s^{\text{DL}}(z)$ again represents the higher-order background components of the saturation mixing ratio. These are the \emph{precipitating DL equations.}

To clarify how rain influences the large-scale flow in this system, let us ignore additional heat sources and turbulent mixing for the moment and substitute the sum of \eqref{PDL_dim_qv} and \eqref{PDL_dim_qr} in the temperature equation \eqref{PDL_dim_temp}, which yields
\begin{equation}
    D_t^g \tilde{\theta}^{\text{DL}} = -\frac{L_{\text{ref}}}{c_{\text{pd}}} \partial_z(V_r q_r^{\text{DL}}).
\end{equation}
Latent heating (cooling) in the moist, precipitating DL regime is therefore \emph{directly proportional to the downward change of the rainfall mixing ratio,} as in section~6 of \citet{smith2017}. However, the internal structure of the moisture dynamics is markedly different in our regime, which is only meant to hold at lower tropospheric levels anyways.
\subsubsection{Matching to the $\text{PQG}_{\text{DL}}$ flow}
In the original DL equations of \citet{klein2022}, the crucial condition for matching to the QG flow could be formulated in terms of the pressure difference over the depth of the diabatic layer. Since this condition - except for minor differences in notation - is formally unchanged in PDL, we only state the result, referring the interested reader to \citet{klein2022} for a detailed derivation: it must hold
\begin{equation}\label{PDL_matching_temp}
    \abs{\int_0^\infty \left(\theta^{(3/2)} - \left.\frac{d\theta_1}{dz}\right\vert_{z=0}\eta' \right) d\eta'}<\infty,
\end{equation}
which is equivalent to the condition that $\theta^{(3/2)}$ approaches its limiting expression sufficiently fast, at a rate faster than $1/\eta$. This will certainly hold if it does so initially \emph{and} all diabatic heating terms vanish at the same rate.

Clearly, the validity of \eqref{PDL_matching_temp} in PDL depends on the evolution of the moist constituents, and the logical next step is to derive appropriate matching conditions for $q_v^{(1/2)}$ and $q_r^{(0/2)}$. Observing that the strength of (negative) latent heating $LD_t^{(0/2)}q_v^{(1/2)}$ in \eqref{PDL_dim_temp} is determined by evaporation $S_{\text{ev}}^{(0/2)}$ in undersaturated air, while no latent heating is felt at leading order in saturated air, the matching condition \eqref{PDL_matching_temp} will certainly be fulfilled if
\begin{equation}\label{PDL_matching_qv}
    q_{\text{vs}}^{(1/2)} - q_v^{(1/2)} = o\left(\frac{1}{\eta}\right) \quad\text{as}\;\eta\rightarrow\infty,
\end{equation}
i.e., if the limit is achieved at a rate faster than $1/\eta$. This condition \emph{will} hold if it does so initially, since no additional sources of moisture appear in \eqref{PDL_dim_qv} (turbulent mixing only redistributes the transported quantity). To be more precise, we observe that, in the source-free non-diffusive case, \eqref{PDL_qvs_pert} implies
\begin{equation}
    D_t^{(0/2)} q_{\text{vs}}^{(1/2)} = ALE e_{s_{00}} D_t^{(0/2)} \theta^{(3/2)},
\end{equation}
which combined with \eqref{PDL_dim_temp} yields
\begin{equation}
    D_t^{(0/2)} \left[q_{\text{vs}}^{(1/2)} - q_v^{(1/2)}\right] = -\left(AL^2E e_{s_{00}} + 1\right) D_t^{(0/2)} q_v^{(1/2)},
\end{equation}
which by \eqref{PDL_dim_qv} is equivalent to
\begin{align}
    D_t^{(0/2)} \left[q_{\text{vs}}^{(1/2)} - q_v^{(1/2)}\right] = & -\left(AL^2E e_{s_{00}} + 1\right) \nonumber \\
    & \times \bar{C}_{\text{ev}} q_r^{(0/2)} \left(q_{\text{vs}}^{(1/2)} - q_v^{(1/2)}\right).
\end{align}
Thus, subsaturation can only decrease along (horizontal) particle trajectories. The addition of external heat sources \emph{can} compensate for this, but the behavior for $\eta \gg 1$ will be unaffected as long as these source terms decay at the appropriate rate.

Now, we check the necessary conditions for matching of the moisture species themselves. To that end, first recall the expansion of $q_v$ in the bulk layer:
\begin{equation}\label{qv_comparison_PQG}
    q_v^{\text{QG}}=q_{\text{vs}0}+\epsilon q_v^{(1)}+o(\epsilon);
\end{equation}
in PDL, we assumed stronger spatiotemporal variations of atmospheric water, translating to
\begin{equation}\label{qv_comparison_PDL}
    q_v^{\text{DL}}=q_{\text{vs}0/2}+\delta q_v^{(1/2)}+o(\delta).
\end{equation}
Taylor expansion of \eqref{qv_comparison_PQG} for $z\ll 1$ and substitution of $z=\delta\eta$ yield
\begin{equation}
    q_v^{\text{QG}}(t,\vect{x},\delta\eta)=q_{\text{vs}0}(0)+ \left.\delta\frac{d q_{\text{vs}0}}{dz}\right\vert_{z=0}\,\eta+o(\delta),
\end{equation}
which must agree with the corresponding terms in \eqref{qv_comparison_PDL} in the limit $\eta\rightarrow\infty$. Thus, matching the water vapor mixing ratio at leading order is equivalent to requiring
\begin{equation}\label{PDL_matching_qv_alt}
    q_v^{(1/2)} - \left.\frac{d q_{\text{vs}0}}{dz}\right\vert_{z=0}\,\eta = o\left(\frac{1}{\eta}\right) \quad\text{as}\;\eta\rightarrow\infty\,.
\end{equation}
However, we already stated \eqref{PDL_matching_qv} as a necessary condition. Next, recall that
\begin{align}\label{eq:qvsDL-FirstOrder}
    q_{\text{vs}}^{(1/2)}&=e^{LA\theta_1(0)}\left[ELA\, T^{(3/2)}+E\,\eta\right] \nonumber \\
    &=e^{LA\theta_1(0)}\left[ELA\,\theta^{(3/2)}-E(LA\Gamma-1)\eta\right] \nonumber \\
    &\sim e^{LA\theta_1(0)} E\left[LA\theta_1'(0)-\Gamma LA+1\right]\eta
\end{align}
in the limit $\eta\rightarrow\infty$, while a straightforward calculation yields
\begin{equation}
    \left.\frac{d q_{\text{vs}0}}{dz}\right\vert_{z=0}=e^{LA\theta_1(0)} E\left[LA\theta_1'(0)-\Gamma LA+1\right],
\end{equation}
which shows that \eqref{PDL_matching_qv_alt} is fulfilled \emph{automatically} when \eqref{PDL_matching_qv} is. This completes our discussion of the matching conditions for the water vapor mixing ratio.

Turning to the constraint \eqref{qc_constraint_dl} for cloud water, we need to impose the following additional conditions for matching to the bulk flow:

In undersaturated PDL regions, the mixing ratio of cloud water must vanish to leading order. Since clouds might still be generated at the bottom of the bulk (QG) layer over time, even in initially undersaturated regions, we need to choose the initial state of the system such that \emph{there is a finite DL height $h$ above which no significant subsaturation occurs.} This is needed in order to make sure that $q_c$ remains continuous across the layers. We thus require that
\begin{equation}\label{qv_sat_dl}
    q_v^{(1/2)}=q_{\text{vs}}^{(1/2)}
\end{equation}
for all $\eta \geq h$. Furthermore, we need to prescribe vanishing diabatic heating, i.e.
\begin{equation}\label{no_heating_dl}
    Q^{(0/2)}=0
\end{equation}
for all $\eta \geq h$, owing to the dependence of $q_{\text{vs}}^{(1/2)}$ on potential temperature perturbations (see Eq.~\eqref{eq:qvsDL-FirstOrder}). If \eqref{qv_sat_dl} holds initially for some finite $h$ and \eqref{no_heating_dl} is prescribed accordingly, this condition will hold for all times. Thus, there is a ``sublayer'' at the top of the DL where clouds that formed in the bulk layer dissolve as we approach strongly undersaturated lower-level air. In other words, $q_c^{(0/2)}$ will continuously tend to zero in said sublayer wherever it is required by \eqref{PDL_dim_qc}, in a manner that otherwise is not specified by the leading-order PDL solution.
    
Physically, this technical condition can be interpreted as follows: clouds that form at lower tropospheric levels extend down into the DL, where their base can be found at finite height. This is consistent with the observed thickness of nimbostratus clouds, which are known to attain a vertical extent of $8-12\text{ km}$ \citep[Chapter 6]{houze2014}. Mathematically, the fact that the decay of $q_c$ in the DL is not fully specified by the model equations strictly speaking constitutes a loss of well-posedness. See section 7 for an outlook on viable alternatives.

Last, but not least, we go over the matching of the respective rain mixing ratios: since this includes only zeroth-order terms, the matching reduces to requiring continuity across the layers,
\begin{equation}\label{PDL_matching_qr}
    q_r^{(0/2)}(t,\vect{x},\eta)\rightarrow q_r^{(0)}(t,\vect{x},0)\quad\text{as}\;\eta\rightarrow\infty.
\end{equation}
In PDL, $q_r$ is determined by the first-order quasi-1D equation
\begin{equation}
    -\partial_\eta q_r^{(0/2)}=-\frac{1}{V_T}\bar{C}_{\text{ev}} \left(q_{\text{vs}}^{(1/2)}-q_v^{(1/2)}\right)^+ q_r^{(0/2)},
\end{equation}
which leads to the representation formula
\begin{align}
    q_r^{(0/2)}(t,\vect{x},\eta)= & q_r^{(0)}(t,\vect{x},0) \nonumber \\
    & \times\exp{\frac{\bar{C}_{\text{ev}}}{V_T}\int_\infty^\eta \left(q_{\text{vs}}^{(1/2)}-q_v^{(1/2)}\right)^+ d\eta'}.
\end{align}
Again, this solution will be valid as long as the integral in the exponent converges, which is already guaranteed by the previously stated matching condition \eqref{PDL_matching_qv}.
\subsection{The Ekman layer: Ekman pumping in PQG-DL-Ekman}
As already stated earlier, the leading-order balances in the Ekman layer are not affected by the incorporation of moist processes: at altitudes $<1\text{ km}$, while clouds can be present, they are most certainly not dense enough over a synoptic region to raise latent heating to a leading-order effect. Thus, the only microphysical mechanism that might be of relevance is evaporation of rain in dry boundary-layer air. Even though this can certainly be very impactful locally, such effects are too weak in the large-scale mean to play a role at leading order. Moreover, recalling that fully developed stratiform clouds typically have their base in the diabatic layer (at altitudes $\sim 1-3\text{ km}$, cf. \citet{xu2019}), we have already incorporated evaporation below cloud base as a strong diabatic effect where it matters the most, i.e., in the PDL equations \eqref{PDL_dim}.

We thus adopt \emph{classical} Ekman theory \citep{pedlosky1987,vallis2017} as the final component of our model, as it was done in \citep{klein2022}. Briefly summarizing the main points, the Ekman solution for the horizontal velocity field provides an expression for the vertical velocity generated in the Ekman layer; matching to the DL solution then allows us to derive the relation
\begin{equation} \label{Ekman_pumping_dim}
    \tilde{w}^{\text{QG}}\vert_{z=0} = \frac{d^{\text{Ek}}}{2}\frac{1}{f} \Delta_\parallel \tilde{\phi^{\text{DL}}}\vert_{z=0}
\end{equation}
in dimensional form, with the \emph{Ekman number} $\text{Ek}$ that is proportional to the turbulent friction coefficient of the flow. This relation now provides the bottom boundary condition for the $\text{PQG}_{\text{DL}}$ pressure perturbation: evaluating the thermodynamic equation \eqref{PQG_preliminary_theta_e} at $z=0$, substituting hydrostatic balance as well as \eqref{Ekman_pumping_dim} and \eqref{PQG_DL_qvs_pert} , redimensionalizing where necessary, we obtain the (dimensional) condition
\begin{align} \label{PQG_DL_dim_bc}
    D_t^g \left(\partial_z \tilde{\phi}^{\text{QG}}\right)\vert_{z=0} =& -g \frac{L_{\text{ref}}}{c_{\text{pd}} T_{\text{ref}}} \frac{1}{B(0)} D_t^g \tilde{M}\vert_{z=0} \nonumber \\
    &- \frac{L_{\text{ref}}/c_{\text{pd}} + B(0)}{B(0)} \frac{d^{\text{Ek}}}{2} \frac{1}{f} \Delta_\parallel \tilde{\phi^{\text{DL}}}\vert_{z=0} \nonumber \\
    & \times \frac{d\bar{\theta}_e}{dz}(0) + Q^{\text{QG}}\vert_{z=0} + \mathcal{D}_\theta^{\text{QG}}\vert_{z=0}
\end{align}
This completes the derivation of our model equations, consisting of the $\text{PQG}_{\text{DL}}$ system \eqref{PQG_DL_dim}-\eqref{PQG_DL_dim_supplement}, the PDL equations \eqref{PDL_dim}-\eqref{PDL_dim_supplement} and the matching conditions \eqref{PDL_matching_temp}, \eqref{PDL_matching_qv}, \eqref{qv_sat_dl}-\eqref{PDL_matching_qr} and \eqref{PQG_DL_dim_bc}.

To illustrate how the bulk flow interacts with the lower-lying boundary layers in a nontechnical manner, Fig. \ref{fig:feedbacks_visual} provides a visual representation of the matching conditions:
\begin{figure}[H]
    \centering
    \includegraphics[width=0.48\textwidth]{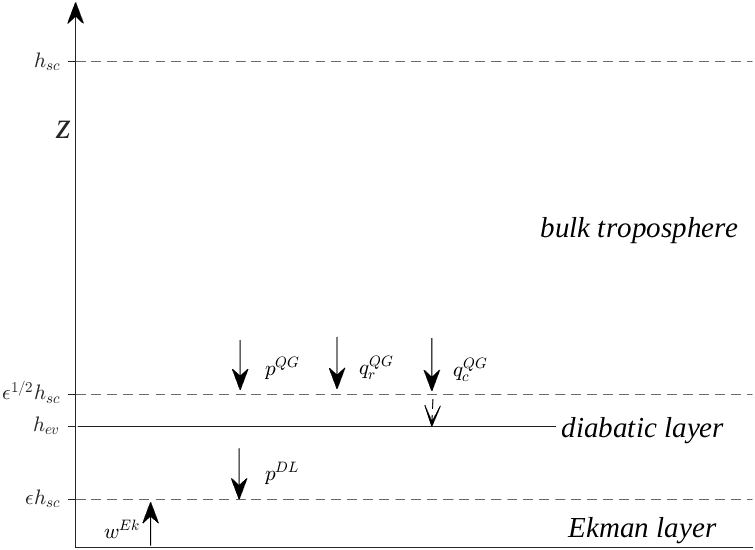}
    \caption{Visual overview of the feedback mechanisms between the individual layers in PQG-DL-Ekman. The vertical arrows are to be interpreted as follows: the value of the variable indicated next to the arrow is imprinted onto the layer it points towards as a consequence of the matching conditions. All interactions of the dry QG-DL-Ekman theory are preserved, with the added effect of rain falling into the DL, where it may partially evaporate. The sublayer between $h_{\text{sc}}$ and $h_{\text{ev}}$ only serves to let clouds dissolve, indicated by the short dashed arrow - note that $h_{\text{ev}}$ can, in principle, also vary in the horizontal.}
    \label{fig:feedbacks_visual}
\end{figure}
\section{Explicit PDL solutions for axisymmetric flow}
We now derive a class of explicit solutions in the precipitating diabatic layer that describe the evolution of idealized vortices, utilizing the scaled equations for notational convenience. It is a useful property of flow in geostrophic balance that certain symmetries annihilate all advection terms in the associated transport equations. In particular, utilizing cylindrical coordinates $(r,\varphi,\eta)$ in the DL and seeking axisymmetric solutions of the form
\begin{equation}
    \vect{U}=\vect{U}(t,r,\eta),
\end{equation}
it immediately follows from \eqref{PDL_geostrophy_1} that
\begin{equation}
    \vect{u}^{(0/2)}=u_\varphi^{(0/2)}\vect{e}_\varphi,
\end{equation}
i.e., the horizontal velocity field is purely azimuthal. Since all transported scalars are independent of $\varphi$, the material derivative $D_t$ reduces to a time derivative and the PDL system without diffusion or external heating reduces to
\begin{subequations}\label{PDL_axisym}
    \begin{align}
        \label{PDL_axisym_temp}
        \partial_t \theta_e^{(0/2)}=&0, \\
        \label{PDL_axisym_qv}
        \partial_t q_v^{(1/2)}=&S_{\text{ev}}^{(0/2)}=\bar{C}_{\text{ev}} T_{00}(q_{\text{vs}}^{(1/2)}-q_v^{(1/2)})^+ q_r^{(0/2)}, \\
        \label{PDL_axisym_qr}
        -\partial_\eta q_r^{(0/2)}=&-\frac{1}{V_T}S_{\text{ev}}^{(0/2)}, \\
        \label{PDL_axisym_qvs}
        q_{\text{vs}}^{(1/2)}=&E e^{AL\theta_1(0)} \nonumber \\
        & \times \left[AL\theta_e^{(3/2)}-AL^2 q_v^{(1/2)}-(AL\Gamma-1)\eta\right],
    \end{align}
\end{subequations}
while pressure and velocity can be diagnostically recovered from the hydrostatic and geostrophic balances, respectively, at any given time. As we shall see shortly, this simplified system can be solved \emph{explicitly} in terms of the solution of the $\text{PQG}_{\text{DL}}$ flow.

\noindent
\emph{Remark:} Axisymmetric solutions are only consistent with the $\text{PQG}_{\text{DL}}$ flow if we neglect the $\beta$-effect. While working on an $f$-plane certainly entails a loss of physical realism, it is common to do so in theoretical models based on QG theory, for example in the famous Eady model for baroclinic instability. We therefore expect the $f$-plane approximation to be sufficient to capture the essentials of the mechanisms that we want to illustrate.
\paragraph{Representation formula for the rain mixing ratio:}
Prescribing an arbitrary initial state that is compatible with the matching conditions, we can write
\begin{align}\label{PDL_axisym_qr_formula}
        q_r^{(0/2)}(t,r,\eta)= & q_r^{(0)}(t,r,0) \nonumber \\
        & \times\exp{\frac{\bar{C}_{\text{ev}}}{V_T}\int_\infty^\eta(q_{\text{vs}}^{(1/2)}-q_v^{(1/2)})d\eta'},
\end{align}
where we recall that $q_r^{(0)}(t,r,0)$ is the rain mixing ratio at the bottom of the QG domain and that the integrand vanishes at \emph{finite} height for all $(t,r)$.
\paragraph{Moist thermodynamics:}
We can reformulate \eqref{PDL_axisym_qv} as an equation for the saturation deficit $q_{\text{vs}}^{(1/2)}-q_v^{(1/2)}$, using \eqref{PDL_axisym_temp} and \eqref{PDL_axisym_qvs}:
\begin{equation}
    \partial_t \theta_e^{(3/2)}=0
\end{equation}
implies
\begin{equation}
    \partial_t q_{\text{vs}}^{(1/2)}=-AEL^2e^{AL \theta_1(0)}\partial_t q_v^{(1/2)} \equiv -C_1 \partial_t q_v^{(1/2)}.
\end{equation}
Further substituting \eqref{PDL_axisym_qr_formula} for $q_r^{(0/2)}$, we arrive at
\begin{align}\label{PDL_axisym_qv_intermediate}
    \partial_t & ( q_{\text{vs}}^{(1/2)}-q_v^{(1/2)}) \nonumber \\
    = & -(C_1+1)V_Tq_r^{(0)}\vert_{z=0} \nonumber \\
    & \times\partial_\eta\left[\exp\left(\frac{\bar{C}_{\text{ev}}T_{00}}{V_T}\int_\infty^\eta(q_{\text{vs}}^{(1/2)}-q_v^{(1/2)})d\eta'\right)\right].
\end{align}
Setting
\begin{equation}
    S(t,r,\eta):=\frac{\bar{C}_{\text{ev}}T_{00}}{V_T}\int_\infty^\eta(\tilde{q}_{\text{vs}}^{\text{DL}}-\tilde{q}_v^{\text{DL}})d\eta',
\end{equation}
we can conveniently rewrite \eqref{PDL_axisym_qv_intermediate} in the form
\begin{equation}
    \partial_t\partial_\eta S=-\bar{C}_{\text{ev}}T_{00}(C_1+1) q_r^{(0)}\vert_{z=0}\partial_\eta e^S,
\end{equation}
which by integration in $\eta$ further simplifies to
\begin{equation}
    \partial_t S=-\bar{C}_{\text{ev}}T_{00}(C_1+1) q_r^{(0)}\vert_{z=0} \left(e^S-1\right).
\end{equation}
The solution to this equation, with initial data $S\vert_{t=0}=S_I(r,\eta)$, is readily obtained in the form
\begin{equation}
    S(t,r,\eta)=-\ln\left[1+\left(e^{-S_I(r,\eta)}-1\right)e^{-R(t,r)}\right],
\end{equation}
where we have set
\begin{equation}
    R(t,r):=\bar{C}_{\text{ev}}T_{00}(C_1+1) \int_0^t q_r^{\text{QG}}(t',r,0)dt'.
\end{equation}
It follows by differentiating in the vertical that
\begin{equation} \label{PDL_axisym_s}
    q_{\text{vs}}^{(1/2)}-q_v^{(1/2)}=:s(t,r,\eta)=s_I(r,\eta)\frac{e^{-S_I(r,\eta)}}{e^{-S_I(r,\eta)}+e^{R(t,r)}-1},
\end{equation}
which shows the decrease of the saturation deficit as a function of the time-integrated rain mass entering the DL from above. Similarly, the rain mixing ratio in the DL results as
\begin{equation}
    q_r^{(0/2)}(t,r,\eta)=\frac{q_r^{(0)}(t,r,0)}{1+\left(e^{-S_I(r,\eta)}-1\right)e^{-R(t,r)}}.
\end{equation}
\paragraph{Expressions for potential temperature, pressure and velocity:}
Denoting the constant-in-time equivalent potential temperature by $E_I^{\text{tot}}(r,\eta)$ and further isolating the unbounded background contributions by introducing
\begin{equation}
    E_I:=E_I^{\text{tot}}-(\theta_1'(0)+L q_{\text{vs}_0}'(0)) \eta,
\end{equation}
we obtain
\begin{equation}
    \theta^{(3/2)}=\frac{1}{C_1+1}\left[E_I(r,\eta)+L s(t,r,\eta)\right]+\theta_1'(0)\eta.
\end{equation}
By \eqref{PDL_pressure_modified}, we get
\begin{equation}
    \phi^{(4/2)}=\int_\infty^\eta \theta^{(3/2)}-\theta_1'(0)\eta'd\eta'+p^{(2)}\vert_{z=0}+\theta_1'(0)\frac{\eta^2}{2},
\end{equation}
which also yields the formula
\begin{align}
   p^{(4/2)}\vert_{\eta=0}= & -\frac{1}{C_1+1}\int_0^\infty E_I(r,\eta')+L s(t,r,\eta')d\eta' \nonumber \\
& +p^{(2)}\vert_{z=0}
\end{align}
for the pressure perturbation at the bottom of the DL.

Evaluating $\nabla_\parallel^\perp p^{(4/2)}$ in polar coordinates, we obtain
\begin{align} \label{PDL_axisym_u}
    & \vect{u}^{(0/2)}(t,r,\eta) \nonumber \\
    = & \left\{\frac{1}{f_0(C_1+1)}\left[\int_\infty^\eta L \partial_r s(t,r,\eta')+\partial_r E_I(r,\eta')d\eta'\right] \right. \nonumber \\
     & \left.+\frac{1}{f_0}\partial_r p^{(2)}\vert_{z=0}(t,r)\right\}\vect{e}_\varphi,
\end{align}
and, by \eqref{PQG_DL_dim_bc}, the Ekman pumping velocity $w^{(1)}\vert_{z=0}(t,r)$ results as
\begin{align} \label{PDL_axisym_wb}
   & w^{(1)}\vert_{z=0}(t,r) = \frac{\sqrt{\text{Ek}}}{2} \frac{1}{f_0} \nonumber \\
   & \times\left\{ \Delta_\parallel \phi^{(2)}\vert_{z=0}(t,r)-\frac{1}{C_1+1} \right. \nonumber \\
    & \left. \quad \times\left[\int_0^\infty L \Delta_\parallel s(t,r,\eta')+\Delta_\parallel E_I(r,\eta')d\eta'\right]\right\}.
\end{align}
Here, $\Delta_\parallel=\partial_{rr}+\frac{1}{r}\partial_r$, owing to axisymmetry.

\section{Numerical solution of the axisymmetric PQG-DL-Ekman system: evolution of a low-level vortex}
We now employ the explicit solutions derived in the preceding section to simplify the numerical treatment of the coupled system in its axisymmetric reduction. While the system thus obtained cannot be expected to reproduce conditions in the real midlatitude atmosphere, both due to the absence of advection and the assumption of (bi)linear phase transitions, it still showcases the nature of interactions across the various layers that will also be present in a more realistic setup. Even with all the simplifications that axisymmetry entails, the $\text{PQG}_{\text{DL}}$ system retains considerable complexity: neglecting diffusion and assuming no external diabatic heating, the dimensional Eqs.~\eqref{PQG_DL_dim} reduce to
\begin{subequations}
    \begin{align}
        \partial_t \text{PV}_e =& 0 \\
        \partial_t \tilde{M} =& B(z)\left[S_{\text{ev}} - S_{\text{cd}}\right] \label{PQG_axisym_M} \\
        \partial_t q_c^{\text{QG}} =& S_{\text{cd}} - S_{\text{ac}} - S_{\text{cr}} \label{PQG_axisym_qc} \\
        -\frac{1}{\bar{\rho}} \partial_z \left(\bar{\rho} V_r q_r^{\text{QG}}\right) =& S_{\text{ac}} + S_{\text{cr}} - S_{\text{ev}} \label{PQG_axisym_qr} \\
        \frac{1}{f}\Delta_\parallel \tilde{\phi}^{\text{QG}} + \frac{f}{\bar{\rho}} & \partial_z \left(\frac{\bar{\rho}}{d\bar{\theta}_e/dz}\frac{B(z)}{L_{\text{ref}}/c_{\text{pd}}+B(z)} \frac{\theta_{\text{ref}}}{g} \partial_z \tilde{\phi}^{\text{QG}}\right) \nonumber \\
         =& \text{PV}_e - \beta y \nonumber \\
        &- \frac{f}{\bar{\rho}} \partial_z \left(\frac{\bar{\rho}}{d\bar{\theta}_e/dz}\frac{L_{\text{ref}}/c_{\text{pd}}}{L_{\text{ref}}/c_{\text{pd}}+B(z)} \tilde{M}\right), \label{PQG_axisym_inversion}
    \end{align}
\end{subequations}
plus hydrostatic and geostrophic relations.

Solving these equations on a cylindrical domain with $z_{\text{max}} = 10\;\text{km}$ and $r_{\text{max}} = 1000\;\text{km}$, other than initial conditions for $\tilde{M}$ and $q_c^{\text{QG}}$, as well as the top boundary condition for $q_r^{\text{QG}}$, we need to specify boundary conditions for the adjusted pressure perturbation $\tilde{\phi}^{\text{QG}}$ in order to obtain a unique solution of \eqref{PQG_axisym_inversion}:
\begin{itemize}
    \item At the bottom of the domain, it is asymptotic matching to the diabatic and Ekman layers that determines the appropriate (time-dependent) boundary condition: by axisymmetry, \eqref{PQG_DL_dim_bc} reduces to
    \begin{align}
        \partial_t \left(\partial_z \tilde{\phi}^{\text{QG}}\vert_{z=0}\right) =& - \frac{g}{\theta_{\text{ref}}} \frac{L_{\text{ref}}/c_{\text{pd}}}{B(0)} \partial_t \tilde{M}\vert_{z=0} \nonumber \\
        &-\frac{L_{\text{ref}}/c_{\text{pd}} + B(0)}{B(0)} \nonumber \\
        & \times \frac{d_{\text{Ek}}}{2} \frac{1}{f} \tilde{w}^{\text{QG}}\vert_{z=0} \frac{d\bar{\theta}_e}{dz}(0),
    \end{align}
    where $d_{\text{Ek}}$ denotes the height of the Ekman layer and $\tilde{w}^{\text{QG}}\vert_{z=0}$ is given by the dimensional equivalent of \eqref{PDL_axisym_wb}. Clearly, an initial condition for $\partial_z \tilde{\phi}^{\text{QG}}\vert_{z=0}$ also needs to be prescribed.
    \item At the top, we impose the rigid lid condition $\tilde{w}^{\text{QG}}\vert_{z=z_{\text{max}}}=0$, which by evaluation of the thermodynamic equation translates to
    \begin{equation}
        \partial_z \tilde{\phi}^{\text{QG}}\vert_{z=z_{\text{max}}} = - \frac{g}{\theta_{\text{ref}}} \frac{L_{\text{ref}}/c_{\text{pd}}}{B(z_{\text{max}})} \tilde{M}\vert_{z=z_{\text{max}}}.
    \end{equation}
    \item At lateral boundaries, we assume homogeneous Neumann boundary conditions for simplicity.
\end{itemize}
Thus, we need to solve a Neumann problem at every time step, which demands that a compatibility condition be satisfied. Here, this will always be the case, as long as the PDL solution is initialized appropriately. Of course, the solution is determined only up to an additive constant, which can be fixed at the initial time.

To showcase how variations in available moisture can shape the flow in this idealized setting, we consider a tropospheric background state exactly at saturation and the following initial perturbation, constructed as a simple harmonic profile, that induces low-level cyclonic flow:

A cloudless free troposphere at rest and a strongly undersaturated DL around the center of the domain, up to a height $h_{\text{ev}} \approx 2.5\;\text{km}$,
    \begin{equation} \label{case1_init_cond}
        s_{\text{DL}}\vert_{t=0} = 0.3 q_{\text{vs}_{\text{ref}}} \frac{1+\cos(\pi r/r_{\text{max}})}{2} H(h_{\text{ev}} - z),
    \end{equation}
where $H$ denotes the Heaviside step function. The constant-in-time equivalent potential temperature perturbation in the DL was set to zero for simplicity. As a further simplification, we assumed zero-$\text{PV}_e$ flow.

As far as the numerical solution method is concerned, the linear elliptic equation \eqref{PQG_axisym_inversion} was solved with standard finite differences on a vertically stretched grid. The evolution Eqs.~\eqref{PQG_axisym_M} and \eqref{PQG_axisym_qc} were discretized with the standard 4th-order Runge-Kutta method, while the diagnostic rainfall Eq.~\eqref{PQG_axisym_qr} was solved by numerically evaluating its integral solution representation at the end of each time step, \emph{after} updated values of $\tilde{M}$, $q_c^{\text{QG}}$ and $\tilde{\phi}^{\text{QG}}$ had been obtained. The a priori undetermined rate constants $\bar{C}_\star$ were pragmatically chosen to fit the order-of-magnitude assumptions underlying our scaling, while the autoconversion threshold $q_{\text{ac}}$ was set to a value reflecting liquid water contents typically observed in stratiform clouds, see Table \ref{tab:rate_constants}:
\begin{table}[H]
\centering
\begin{adjustbox}{width=0.48\textwidth}
\begin{tabular}{l l l l}
\hline
    $\bar{C}_{\text{ev}} \frac{\bar{p}}{\bar{\rho}}$ & 0.1 & $s^{-1}$ & Evaporation rate, multiplied by constant factor $\bar{p} / \bar{\rho}$ \\
    $\bar{C}_{\text{cd}}$ & 0.01 & $s^{-1}$ & Condensation rate \\
    $\bar{C}_{\text{cn}} q_{\text{cn}}$ & 0.1 & $s^{-1}$ & Nucleation rate, multiplied by density of condensation kernels $q_{\text{cn}}$ (assumed constant) \\
    $\bar{C}_{\text{ac}}$ & $10^{-5}$ & $s^{-1}$ & Autoconversion rate \\
    $q_{\text{ac}}$ & $4 \times 10^{-4}$ & 1 & Autoconversion threshold  \\
    \hline
\end{tabular}
\end{adjustbox}
\caption{Rate constants in the bulk microphysics closures. Notice that $\bar{C}_{\text{cn}} q_{\text{cn}}$ was set to a value that keeps supersaturation one order of magnitude smaller than subsaturation, consistent with observational data.}
\label{tab:rate_constants}
\end{table}
The code used to generate the figures in this section was written in MATLAB; the corresponding author will be happy to provide it upon personal request.
\subsection*{Tropospheric dynamics generated by a low-level moisture trough}
\begin{figure}[H]
    \centering
    \includegraphics[width=0.24\textwidth]{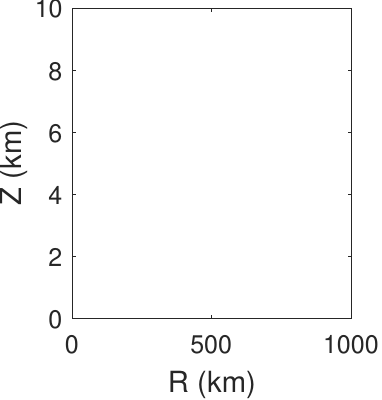}
    \includegraphics[width=0.24\textwidth]{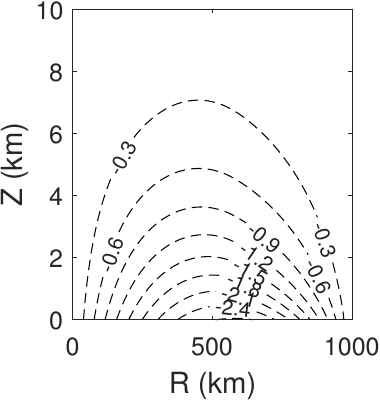}
    \includegraphics[width=0.24\textwidth]{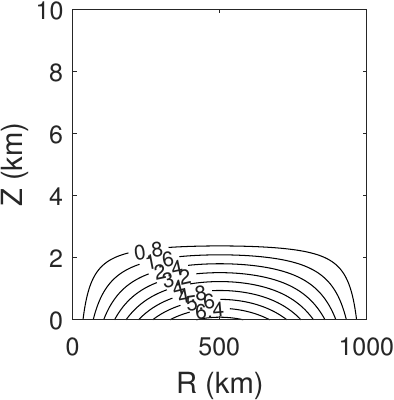}
    \includegraphics[width=0.24\textwidth]{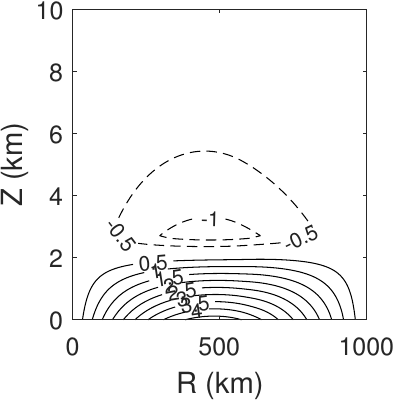}
    \includegraphics[width=0.24\textwidth]{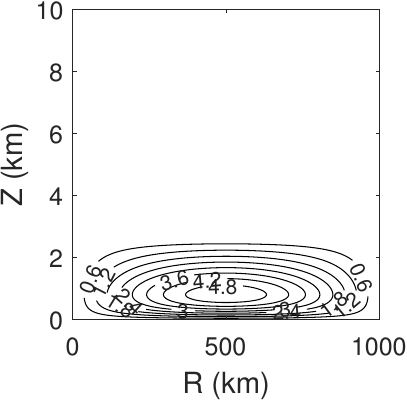}
    \includegraphics[width=0.24\textwidth]{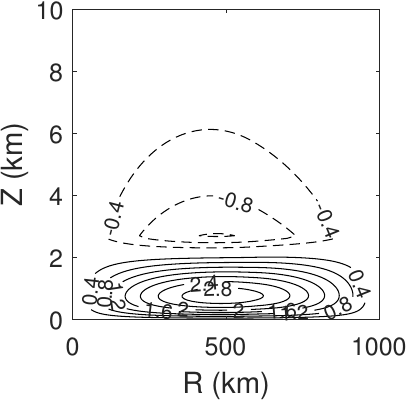}
    \caption{Top to bottom: PQG, composite PQG-DL and full PQG-DL-Ekman profiles of the azimuthal velocity at the initial time (left) and at $t=12$ h (right). Notice the effect of surface friction in the full profile and the emergence of anticyclonic flow in the PQG solution component.}
    \label{fig:u_0-12}
\end{figure}

In the scenario described above, with \eqref{case1_init_cond} as the only nonzero initial condition, the tropospheric flow is wholly driven by updrafts produced in the Ekman layer, which in turn are effected by the horizontal moisture differential in the DL. The latter sets up a cyclonic flow in the dry sublayer that is brought to rest at the bottom by surface friction, while over time a corresponding \emph{anticyclonic} flow develops in the middle and upper troposphere (Fig.~\ref{fig:u_0-12}). This flow gradually intensifies and weakens the low-level cyclonic motion.
This behavior is largely analogous to that observed in the simulation of a heat low in \citep{klein2022} - in this application of the \emph{dry} QG-DL-Ekman theory, however, an external heat source (parameterizing small-scale convection) was responsible for said development, which here arises directly from the synoptic-scale moisture distribution. The conspicuous edges arise from the cutoff in \eqref{case1_init_cond} at $z = h_{\text{ev}}$.

\begin{figure}[H]
    \centering
    \includegraphics[width=0.24\textwidth]{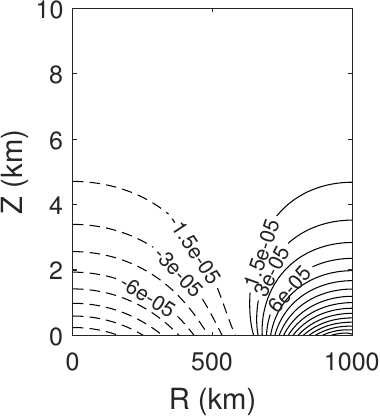}
    \includegraphics[width=0.24\textwidth]{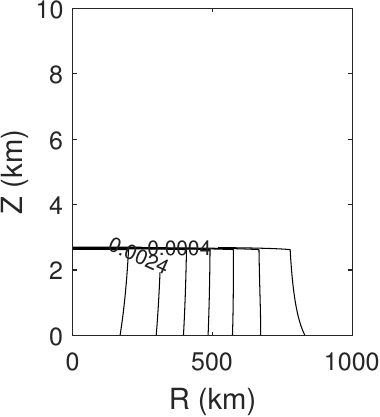}
    \includegraphics[width=0.24\textwidth]{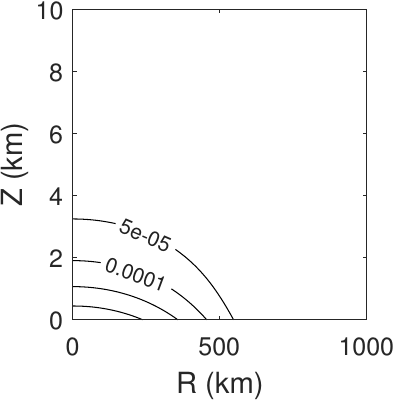}
    \includegraphics[width=0.24\textwidth]{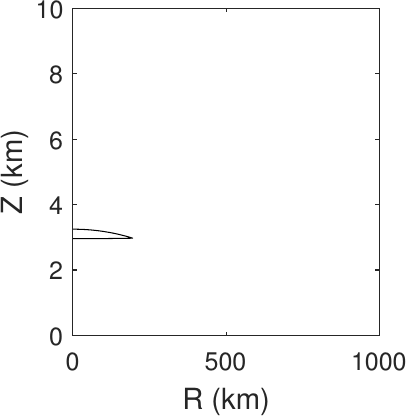}
    \caption{Left to right, top row: absolute subsaturation $\tilde{q}_{\text{vs}} - \tilde{q}_v$ in the free troposphere and in the PQG-DL composite solution at $t=6$ h. On the left, a saturated inner and an unsaturated outer region emerge. In the composite solution on the right, where DL subsaturation dominates, only the lowest and the highest value are displayed for readability; here, contours coalesce at $z = h_{\text{ev}}$ due to the cutoff prescribed in \eqref{case1_init_cond}. Left to right, bottom row: cloud water mixing ratio in the free troposphere and in the PQG-DL composite solution at $t=6$ h. In the latter, the cloud layer is still very thin.}
    \label{fig:beta_qc_6}
\end{figure}

\begin{figure}[H]
    \centering
    \includegraphics[width=0.24\textwidth]{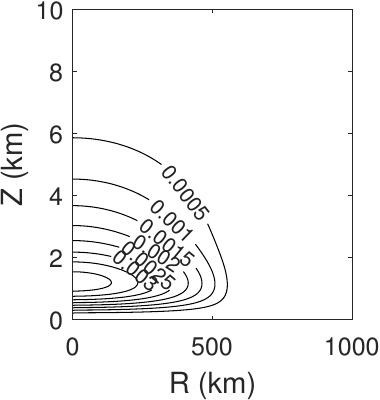}
    \includegraphics[width=0.24\textwidth]{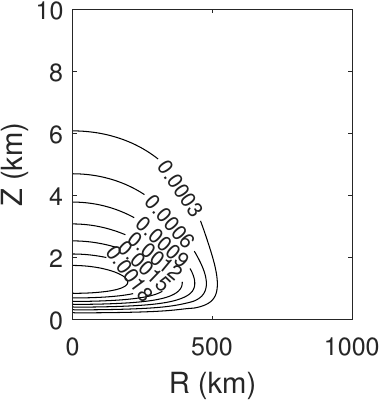}
    \caption{Left to right: composite PQG(-DL)-Ekman vertical velocity at $t=6$ h and $t=24$ h. Only updrafts due to Ekman pumping are present. These updrafts decrease in strength over time.}
    \label{fig:wQGEk_6_24}
\end{figure}

\begin{figure}[H]
    \centering
    \includegraphics[width=0.24\textwidth]{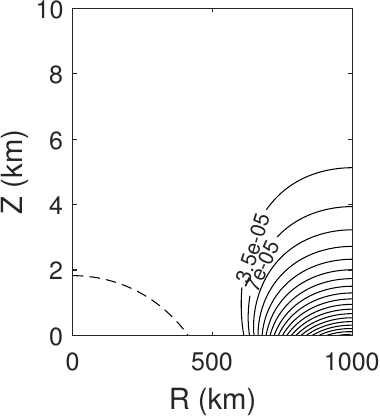}
    \includegraphics[width=0.24\textwidth]{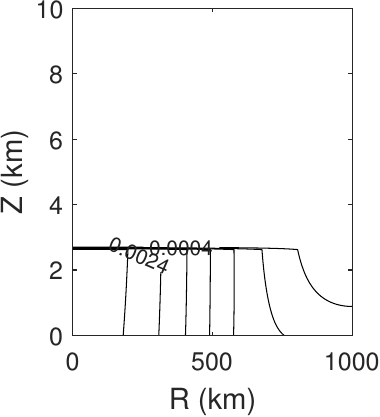}
    \includegraphics[width=0.24\textwidth]{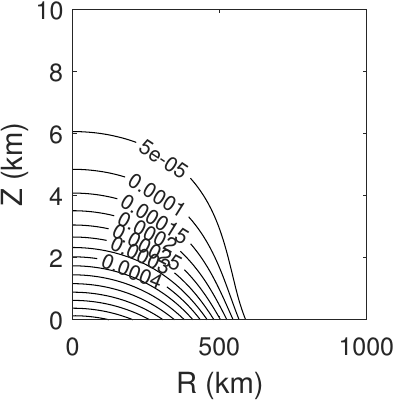}
    \includegraphics[width=0.24\textwidth]{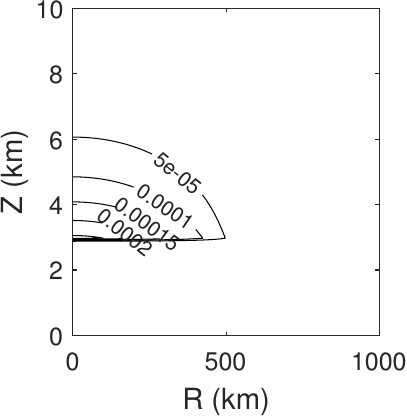}
    \caption{Left to right, top row: free tropospheric and composite subsaturation at $t=24$ h. In the composite profile, the $\text{PQG}_{\text{DL}}$ contribution has altered the surface moisture distribution close to the lateral boundary. Left to right, bottom row: free tropospheric and composite cloud water at $t=24$ h. In the saturated region, clouds have grown much denser; in the composite profile, contours bunch up close to $h_{\text{ev}}$ because the cloud dissolves in a very thin sublayer.}
    \label{fig:beta_qc_24}
\end{figure}

\begin{figure}[H]
    \centering
    \includegraphics[width=0.24\textwidth]{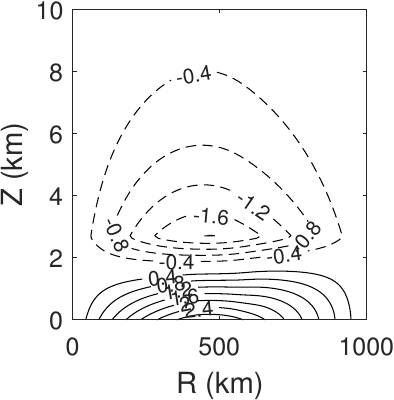}
    \includegraphics[width=0.24\textwidth]{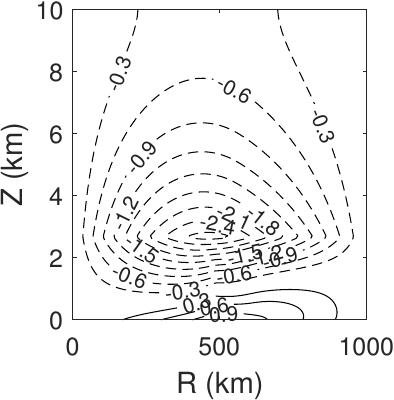}
    \includegraphics[width=0.24\textwidth]{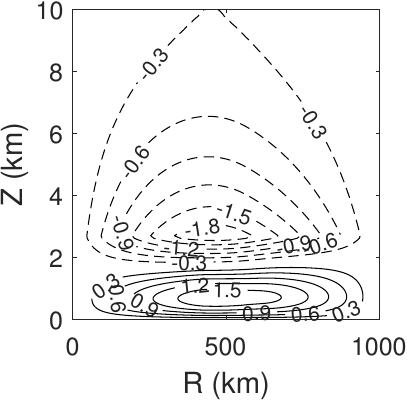}
    \includegraphics[width=0.24\textwidth]{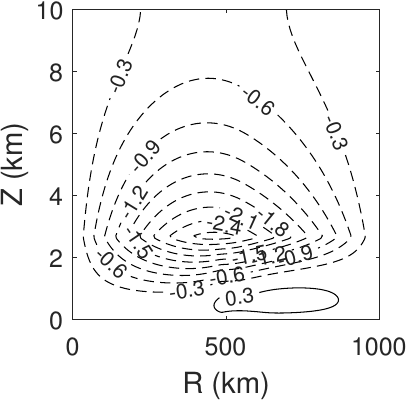}
    \caption{Top row: PQG-DL composite azimuthal velocity at $t=24$ h (left) and at the final time $t=48$ h (right). The isolines of the low-level cyclonic flow start to bend slightly downward below the saturated region, and at the final time, continuous evaporation of drizzle generated in the free troposphere has accelerated the dissolution of the surface cyclone below the cloudy region. Bottom row: PQG-DL-Ekman composite azimuthal velocity at $t=24$ h (left) and at the final time $t=48$ h (right). Apart from the weakening of the cyclonic flow, the same trend as in the PQG-DL profile can be observed. At the final time, the tendency of evaporative cooling to dissolve the surface cyclone is even more apparent.}
    \label{fig:uQGDL_uQGDLEk_24_48}
\end{figure}

As far as moisture in the free troposphere is concerned, an immediate subdivision into a saturated inner and an undersaturated outer region can be observed. Air in the outer region progressively dries out over time, while it stays close to the saturation threshold in the inner region, where a cloud layer forms. These developments are shown in Fig.~\ref{fig:beta_qc_6}. In this part of the domain, upwelling due to DL-Ekman pumping is significant, while no discernible vertical motion is generated in the outer region (Fig.~\ref{fig:wQGEk_6_24}, left). Here and in the following, the composite profile of the cloud water mixing ratio only represents a qualitatively accurate approximation: as discussed in subsection~4b, the manner in which $q_c$ tends to zero as we approach the dry subregion of the DL is not fully specified by the model output. Still, one can prescribe a two-component partition of unity $\chi_i = \chi_i(z)$ ($i = 1,2$) in the vertical and write
\begin{equation}
    q_c^{\text{DL}} = \chi_1 \cdot 0 +\chi_2 q_c^{\text{QG}},
\end{equation}
where $\chi_1,\chi_2 \geq 0$, $\chi_1 + \chi_2 \equiv 1$, $\chi_2 = 1$ in the QG layer and $\chi_1 = 1$ below $h_{\text{ev}}$. This method of representing a solution that is undetermined in some transition region has been used in the context of formal asymptotic analysis, e.g., by \citet{tordeux2006}. The composite subsaturation profile also constitutes a rougher approximation than the other solution components, since the PDL solution does not incorporate the $O(\epsilon^3)$ perturbations of $q_{\text{vs}}^{\text{DL}}-q_v^{\text{DL}}$ that would formally match the corresponding quantities in the QG layer.
At later times ($t=24$ h), updrafts weaken considerably (Fig.~\ref{fig:wQGEk_6_24}, right). As far as the moisture distribution is concerned, the initial trend continues, as can be seen in Fig.~\ref{fig:beta_qc_24}. Here, $q_c$ has already crossed the autoconversion threshold, which was set at $q_{\text{ac}} = 0.4\;\text{g/kg}$.
Remarkably, while the rain mixing ratio $q_r$ is almost negligibly small (and therefore not shown), its impact can nevertheless be seen in the composite azimuthal velocity at $t=24$ h: since DL subsaturation is weakened by rain entering the layer from above, according to \eqref{PDL_axisym_s}, and the velocity field in the diabatic layer is shaped by the horizontal moisture distribution as expressed in \eqref{PDL_axisym_u}, new structures - albeit still subtle - start to emerge at low levels. At the final time, $t=48$ h, no significant changes in the evolution of the tropospheric moisture distribution can be observed, other than the dampening of cloud growth beyond the autoconversion threshold due to continuous conversion of cloud droplets into raindrops. The time-integrated rain mass entering the DL from above, however, has reshaped the low-level flow almost completely, as illustrated by Fig.~\ref{fig:uQGDL_uQGDLEk_24_48}.
In summary, the numerical sample solution illustrates how
\begin{enumerate}
    \item[(i)] a moisture anomaly in the diabatic layer can generate long-lasting dynamical disturbances across the whole troposphere from a stationary state and
    \item[(ii)] rain generation in the bulk troposphere affects the flow closer to the surface.
\end{enumerate}
\section{Outlook on model variants}
\subsection{PQG-DL-Ekman in the regime $\alpha=0$}
In the present study, we have chosen to investigate only one out of the two distinguished limits in Table \ref{tab:regimes_moist}. As explained at the beginning of section 3, our regime of choice ($\alpha=1$) was adopted mainly due to its compatibility with the PQG model family; in particular, it led to leading-order balances that preserved the functional form of hydrostatic balance in QG theory for dry air. If one adopts the formally consistent regime $\alpha=0$ instead, leaving all other scaling assumptions unchanged, the formal derivation yields a modified hydrostatic relationship of the following form (in both the free troposphere and the DL):
\begin{equation}
   \partial_z \tilde{\phi} = g \left( \frac{\tilde{\theta}}{\theta_{\text{ref}}} + \frac{R_v}{R_d} \tilde{q}_v \right),
\end{equation}
where, as in the rest of the article, $\tilde{\phi}$ denotes the (vertically shifted) pressure perturbation scaled by the background density, $\tilde{\theta}$ the potential temperature perturbation and $\tilde{q}_v$ perturbations of the water vapor content about the moist background. This constitutes a significant departure from all moist extensions of QG theory considered thus far in the literature, since it incorporates water vapor buoyancy at leading order. There are, however, empirical studies indicating a substantial influence of vapor buoyancy on tropospheric (thermo)dynamics at least in (sub)tropical regions, see, e.g., \citet{yang2022}. For this reason, we intend to devote a follow-up article to a close investigation of both pure PQG dynamics and the triple-deck theory developed here in this alternative regime.
\subsection{A precipitating DL with denser clouds}
In deriving the PDL system \eqref{PDL_dim}, we were guided by the observation that CLWCs in stratiform clouds typically do not significantly exceed $\sim 1$ g/kg, and therefore did not assume an asymptotic increase in cloud density in the DL relative to the free troposphere. This led to a model in which evaporation of rain in undersaturated air emerged as the sole dominant moist process in the DL. We should not, however, dismiss the possibility that \emph{local} thickening of the cloud cover at lower levels might make its presence felt in the large-scale dynamics. Therefore, we also plan to investigate a variant of PQG-DL-Ekman based on the scaling
\begin{equation}
    q_c^{\text{DL}} = O(\epsilon^{5/2})
\end{equation}
in future work.
\section{Conclusions}
A new reduced model describing moist synoptic-scale dynamics in the midlatitudes was derived. The derivation of said model, the PQG-DL-Ekman theory, builds on foundations established in a fair number of earlier asymptotic modeling studies: first, the general framework of \citet{klein2010} for model hierarchies in atmospheric flows; then, the studies of \citet{klein2006} and \citet{hittmeir2018} on moist convection, which systematically incorporated the Clausius-Clapeyron relation and investigated possible extensions of the dry air distinguished limit of \citet{klein2010} to include the thermodynamics of moist air; the PQG model family of \citet{smith2017}, which was later shown by \citet{baeumer2023} to be derivable from the \citet{klein2010} distinguished limit and finally the QG-DL-Ekman theory of \citet{klein2022}.

Beyond the formal derivation and the description of the resulting triple-deck boundary layer theory, a simplified axisymmetric version of the model was studied in more detail to elucidate its core features: in the precipitating DL, this simplification enabled the explicit representation of the system's solution in terms of the required data. With those explicit solutions in hand, numerical solutions of the coupled (axisymmetric) PQG-DL-Ekman system could also be computed by standard methods. The initial state was chosen to illustrate the impact of nonuniform moisture distributions at low tropospheric levels, in particular the potential of a dry DL sublayer to shape the whole tropospheric flow through the effects of moist processes.

Concerning directions for future research, the various modeling alternatives mentioned in section~7 constitute an obvious starting point. Beyond those, we have already begun to investigate the formation of surface fronts in the precipitating DL; rigorous PDE studies of the DL equations are also in the works. We further plan to explore in more depth the relationship between models utilizing bulk microphysics closures - like the present one - and those that model phase changes by switching functions, such as \citet{smith2017} and most subsequent studies of the PQG equations. Such work should prove to be helpful in improving our understanding of the link between cloud microphysics and large-scale atmospheric dynamics. Finally, we want to point to the long-term goal of our asymptotic modeling endeavors, which is a unified multiscale description of cloud formation and precipitation in the earth's atmosphere, from the convective cloud towers of \citet{hittmeir2018} to the synoptic-scale description of moist dynamics of \citet{smith2017,baeumer2023} and the present article.

\acknowledgments
This research was funded in whole or in part by the Austrian Science Fund (FWF) 10.55776/F65. For open-access purposes, the authors have applied a CC BY public copyright license to any author-accepted manuscript version arising from this submission. R.K.~acknowledges support by Deutsche Forschungsgemeinschaft through Grant CRC 1114 ``Scaling Cascades in Complex Systems'', Project Number 235221301, Project C06 ``Multi-scale structure of atmospheric vortices'' and Grant FOR 5528 ``Mathematical Study of Geophysical Flow Models:
Analysis and Computation'', Project No.~500072446, Project~2 ``Scale Analysis and Asymptotic Reduced Models for the Atmosphere''. Both authors thank the Wolfgang Pauli Institute Vienna for all kinds of support, e.g., the Pauli fellowship for R.K. \\
Further we acknowledge continous scientific discussions with N.J. Mauser, and valuable help with the numerical implementation by H.P. Stimming.

\datastatement
The paper presents theoretical work and all derivations should be described in sufficient detail. No further data are required to reproduce the findings.

\appendix

\section*{The fast condensation limit}
Let us first clarify the notion of ``fast condensation'', which corresponds to the following implicit definition of the condensation term:
\begin{align}\label{condensation_implicit}
    \left\{\begin{array}{ll}
        q_v=q_{\text{vs}},\ \ \ q_c=q_t-q_{\text{vs}}-q_r & \text{in saturated air} \\
        q_v<q_{\text{vs}},\ \ \ q_c=0 & \text{in undersaturated air,}
    \end{array}\right.
\end{align}
where $q_t$ denotes the total water content. This definition effectively reduces the number of prognostic variables by one, since $q_c$ can be calculated explicitly from $q_t$ and $q_r$ at any given time (and $q_{\text{vs}}$, which in our model is a function of the potential temperature perturbation, but often is assumed to be given as a function of $z$):
\begin{equation}
    q_c = \max \left(q_t-q_{\text{vs}}-q_r, 0\right).
\end{equation}
By the same token, we have
\begin{equation} \label{implicit_qv}
    q_v = \min \left(q_{\text{vs}}, q_t - q_r\right).
\end{equation}
On a formal level, this alternative can be recovered systematically simply by rescaling the respective nucleation and condensation rates appropriately: instead of assuming continuous reparameterization, as described in section~3b, we straightforwardly identify both $C_{\text{cn}}$ and $C_{\text{cd}}$ as asymptotically fast with respect to the chosen timescale and make the ansatz
\begin{equation}
    C_{\text{cn}} t_{\text{ref}} \sim C_{\text{cd}} t_{\text{ref}} \sim \epsilon^{-n},
\end{equation}
for some $n \gg 1$. The leading-order balance in the scaled moisture equations then results as
\begin{equation}
    S_{\text{cd}}^{(0)} = C_{\text{cn}} (q_v^{(1)} - q_{\text{vs}}^{(1)})^+ q_{\text{cn}}^{(0)} + C_{\text{cd}} (q_v^{(1)} - q_{\text{vs}}^{(1)}) q_c^{(0)} = 0,
\end{equation}
which, due to the nonnegativity of $q_c^{(0)}$ and $q_{\text{cn}}^{(0)}$, immediately enforces \eqref{condensation_implicit}. We thus obtain a formal derivation of the alternative that was prescribed by \citet{wetzel2019} from the outset. Notice that our model also differs from that of Wetzel and coauthors in the scaling of the terminal fall velocity.

Now, let us sketch how the anelastic PQG equations of \citet{wetzel2019} can be derived from the leading-order balances on the synoptic timescale. In doing so, we first need to take into consideration the deviating scalings in the cited article: Wetzel et al. assumed a terminal rainfall velocity comparable to the vertical reference velocity,
\begin{equation}
    V_r \sim w_{\text{ref}},
\end{equation}
and consequently a rain mixing ratio of the same order of magnitude as the overall moisture anomaly,
\begin{equation}
    q_r^{\text{QG}} \sim \tilde{q}_v^{\text{QG}}.
\end{equation}
Finally, Wetzel and coauthors assumed a saturation mixing ratio that was given as a background profile at \emph{all} orders, which simplifies the further treatment. With these scalings in place and proceeding purely formally from a regular asymptotic expansion ansatz, the leading-order equations for the moist thermodynamics read in our notation
\begin{subequations} \label{PQG_Wetzel_prelim}
    \begin{align}
        D_t^{(0)} \theta_e^{(2)} + w^{(1)} \frac{d \theta_{e_1}}{dz} &= 0 \\
        D_t^{(0)} q_v^{(1)} +w^{(1)} \frac{d q_{\text{vs}_0}}{dz} &= S_{\text{ev}}^{(0)} - S_{\text{cd}}^{(n)} \label{PQG_Wetzel_prelim_qv} \\
        D_t^{(0)} q_c^{(0)} &=  S_{\text{cd}}^{(n)} - S_{\text{ac}}^{(0)} - S_{\text{cr}}^{(0)} \\
        D_t^{(0)} q_r^{(0)} - \frac{1}{\rho_0} \partial_z \left(\rho_0 V_T q_r^{(0)}\right) &=  S_{\text{ac}}^{(0)} + S_{\text{cr}}^{(0)} - S_{\text{ev}}^{(0)}, \label{PQG_Wetzel_prelim_qr}
    \end{align}
\end{subequations}
where we assumed a non-diffusive regime. Since, upon taking the fast condensation limit, $q_v^{(1)}$ and $q_c{(0)}$ become functions of $q_t^{(1)} = q_v^{(1)} + q_c^{(0)} + q_r^{(0)}$ and $q_r^{(0)}$, we rewrite \eqref{PQG_Wetzel_prelim} accordingly in a reduced form, taking the sum of \eqref{PQG_Wetzel_prelim_qv}-\eqref{PQG_Wetzel_prelim_qr} to obtain an equation for $q_t^{(1)}$:
\begin{subequations} \label{PQG_Wetzel}
    \begin{align}
        D_t^{(0)} \theta_e^{(2)} + w^{(1)} \frac{d \theta_{e_1}}{dz} &= 0 \\
        D_t^{(0)} q_t^{(1)} &= \frac{1}{\rho_0} \partial_z \left(\rho_0 V_T q_r^{(0)}\right) \\
        D_t^{(0)} q_r^{(0)} - \frac{1}{\rho_0} \partial_z \left(\rho_0 V_T q_r^{(0)}\right) &=  S_{\text{ac}}^{(0)} + S_{\text{cr}}^{(0)} - S_{\text{ev}}^{(0)}.
    \end{align}
\end{subequations}
These equations - together with geostrophic and hydrostatic balances, as well as the QG vorticity equation - form a closed system in the limit $\tau \rightarrow \infty$.

Concerning the details of the derivation of a suitable PV formulation, we refer the reader to \citet{wetzel2019}. The key point here is that \eqref{implicit_qv} can be written in the form
\begin{equation}
    q_v = H_u (q_t - q_r) + H_s q_{\text{vs}},
\end{equation}
where $H_u$ and $H_s$ denote indicator functions of undersaturated and saturated states, respectively; however, when rewriting \eqref{PQG_Wetzel} in terms of balanced quantities, the local saturation state cannot be determined without knowledge of $\theta$, which is part of the \emph{output} of the PV-and-moisture inversion equation. The nonlinearity of said equation (and the appearance of a switch in the buoyancy frequency therein) can thus be seen as a consequence of the combination of a moist background with fast microphysics.

\noindent
\emph{Remark:} The analytical and numerical investigation of piecewise elliptic operators with switches at the phase interface certainly poses numerous challenges. We do want to mention, however, that progress on the analysis front has been made recently: \citet{remond-tiedrez2024} have established the well-posedness of weak solutions of the original PQG PV-M inversion equation of \citet{smith2017}.

\bibliographystyle{ametsocV6}
\bibliography{references}

\end{document}